\documentclass[11pt]{article}
\usepackage{titlesec}
\usepackage[toc]{appendix} 
\pdfoutput=1
\usepackage{tikz,amsmath}
\usetikzlibrary{calc}
\usepackage{bbm}
\usepackage{feynman}
\usepackage{xcolor}
\usepackage{dsfont}
\usepackage{subfigure}
\usepackage{authblk}
\usepackage{hyperref}
\usepackage{microtype}
\usepackage{caption}
\usepackage{amsmath}
\usepackage{float}
\usepackage{graphics}
\usepackage{psfrag}
\usepackage{color}
\usepackage{slashed}
\usepackage{subfigure}
\usepackage{graphicx}
\usepackage{array,multirow,axodraw}
\usepackage{amsfonts}
\usepackage{amssymb}
\usepackage{mathtools}
\usepackage{amsmath}
\usepackage{amsfonts}
\usepackage{mathrsfs}
\usepackage{multirow}
\usepackage{feynmp}
\usepackage{array}
\usepackage{textcomp} 
\usepackage{phaistos} 
\usepackage[utf8]{inputenc}
\usepackage{pifont} 
\usepackage{braket}
\usepackage{subfigure}
\usepackage{multirow}
\usepackage{dsfont}
\usepackage[normalem]{ulem}

\definecolor{darkgreen}{cmyk}{1,0,1,0.4}

\renewcommand{\baselinestretch}{1.2}
\renewcommand{\arraystretch}{1.5}
\def\beq{\begin{equation}}
\def\eeq{\end{equation}}
\def\barr{\begin{array}}
\def\earr{\end{array}}

\newcommand{\be}{\begin{equation}}
\newcommand{\ee}{\end{equation}}
\newcommand{\bea}{\begin{eqnarray}}
\newcommand{\eea}{\end{eqnarray}}

\newcommand{\bi}{\begin{itemize}}
	\newcommand{\ei}{\end{itemize}}
\newcommand{\ord}{{\cal O}}
\newcommand{\nc}{\newcommand}
\nc{\LL}{L}
\nc{\vv}{\tilde{v}}
\nc{\GG}{\tilde{G}}
\begin{document}
\title{Taming the $\epsilon_K$ in Little Randall Sundrum Models }
\author[1,2]{Giancarlo D'Ambrosio}
\affil[1]{INFN-Sezione di Napoli,   Complesso Universitario di Monte S. Angelo, Via Cintia Edificio 6, 80126 Napoli, Italy}
\affil[2]{ Centre for High Energy Physics, Indian Institute of Science, C. V. Raman Avenue, Bangalore 560012, India}
\author[2]{Mathew Thomas Arun}
\author[2,1]{Ashwani Kushwaha}
\author[2]{Sudhir K. Vempati}

\maketitle
\begin{abstract}
The Randall Sundrum (RS) models receive significant constraints
from the neutral Kaon system.  The CP violating observable $\epsilon_K$, in Randall Sundrum scenario, requires the lightest KK gluon to be heavier than $\sim$ 24 TeV. The constraint is even  stronger in the Little Randall Sundrum models (LRS), $\gtrsim$ 32 TeV. The LRS models are motivated for their possible visibility at the Large Hardon Collider (LHC).  We show that the  stringent constraints from K-physics  can be relaxed in the LRS models, in the presence of the Brane Localised Kinetic Terms (BLKT). In particular, for a range of values, a UV BLKT could significantly modify the lightest KK gluon wave function such that the limit can reduces to 5 TeV. We also show that such a relaxation of the constraints can also be achieved by imposing flavour symmetries {\`a la } Minimal Flavour Protection.

\end{abstract}

\newpage
\section{Introduction}
Flavour observables have been at the forefront of constraining physics beyond Standard Models. Precision observations of rare decays with high experimental accuracy have put significant constraints on new physics. The rare $\mu \to e+ \gamma$ decay with a branching fraction limit of the $ \sim \mathcal{O}(10^{-13})$ is one such example which puts severe constraints on various supersymmetric and extra-dimensional theories. In the hadronic sector the neutral K meson system puts the strongest constraint through observables such as $\Delta m_K$, $\epsilon_K$ and ${\epsilon'/ \epsilon}$. Of all the three, $\epsilon_K$ remains the strongest constraint (For recent reviews, please see \cite{Beall:1981ze,Buchalla:1995vs,Buras:1998raa,Isidori:2010kg,Kiers:2002cz,Gori:2016lga,Zupan:2019uoi}). Randall Sundrum \cite{Randall_1999_1,Randall_1999_2} models too have been subjected to this scrutiny by the neutral K meson system \cite{Agashe:2003zs,Agashe:2004cp}.
In the Randall Sundrum (RS)  model the dominant new contribution to the $K^0-\bar{K}^0$ mixing comes through the exchange of lowest KK gluon ($g^{(1)}$) at the tree level. To be compatible with the experiments, the lower limit on the mass of the $g^{(1)}$ should be of the $\mathcal{O}( 24~\text{TeV})$ . This limit is very stringent and rules out any possibility of producing $g^{(1)}$ at the LHC. 

The Little Randall Sundrum (Little RS) model \cite{Davoudiasl:2008hx,Davoudiasl_2010} is a version of the RS model where the UV scale is reduced to $\Lambda \sim$~100 TeV from the 
Planck scale. The attractive features of this model include enhanced signals at LHC, suppressed contributions to electro-weak precision observables \cite{Davoudiasl:2008hx} etc. However, the constraints from $\epsilon_K$ turn for the worse, they become more stringent \cite{Bauer:2008xb}. A direct comparison between RS and Little RS would be misleading as the same bulk mass parameters would not fit fermion masses and mixing. A more accurate comparison would be to fit the fermion masses in both models and compare the likely lower bound on the mass of $g^{(1)}$.  It is shown that the limit is $\sim \mathcal{O}( 32~\text{TeV})$, an increase of about $30\%$. Thus it would be interesting to see for mechanisms which could soften the constraints on Little RS.

Brane Localised Kinetic Terms (BLKTs) have been first discussed in the context of gravity \cite{Dvali:2000hr} and for counterterms arising from loop corrections to the bulk field propagators \cite{Georgi:2000ks} in extra dimensions compactified on an orbifold. Later, the idea was successfully applied to gauge theories \cite{Carena:2002dz,Carena:2003fx} to solve the ``little" hierarchy problem in the electro-weak sector in the RS model. Flavour physics, in particular the Kaon system, in Little RS has a similar pathology. Hence, it would be natural to ask whether BLKT could be useful in addressing the strong constraints in the model. We show that for a range of values for the BLKTs, the lower limits on $g^{(1)}$ can be softened significantly.  
Another interesting approach would be the application of flavour symmetries like U(2) or U(3). In the context of 
RS models, this has been used by \cite{Santiago:2008vq}, as Minimal Flavour Protection (MFP). We apply this paradigm to the
Little RS model and show that the bounds on $g^{(1)}$ can become as low as 5 TeV. 

The paper is organised as follows. In the next section we review the Little RS model, and along the way set up the notation used in the rest of the paper. In Sec.3 we will detail the computation of $\epsilon_K$ in the little RS model. In the next section, Sec.4, we discuss the Little RS model in the presence of Brane Localised Kinetic Terms and compute the constraints from $\epsilon_K$. Sec.5 is devoted to imposition of flavour symmetries U(2) and U(3) and their implications. We end with a small summary and outlook.

\section{Recap of the little  RS model}
 To make the present paper self contained, we briefly recap the Little RS
 model in this section.
 The extra-dimensional set up is the same as the Randall Sundrum model \cite{Randall_1999_1} with the background metric defined on a $M_4 \times S^1/Z_2$ orbifold with a negative bulk cosmological constant. This geometry is defined by the line element
\begin{eqnarray}
ds^{2}= g^{MN}g_{MN} = e^{-2ky}\eta_{\mu \nu}dx^{\mu}dx^{\nu}+dy^{2}
\label{geometry}
\end{eqnarray}
where M,N are 5 dimensional space-time indices, $\eta_{\mu \nu}=diag(-1,+1,+1,+1)$ and $0\le y\le L$. The warp factor $k$, is set such that $k L \sim 7$ with the fundamental scale $M_5 \sim {\cal O}(10^3 \ \text{TeV})$ \cite{Davoudiasl:2008hx} much lower than the UV scale in the RS case.

We assume the bulk to be populated by gauge fields and fermions transforming under the adjoint representation and fundamental representations of the Standard Model gauge group $SU(3)\times SU(2)_{L}\times U(1)_{Y}$ respectively \cite{Agashe:2003zs}. The Higgs field, which transforms as a doublet under the weak gauge group, is assumed to be localized on the IR brane. This stabilises the Higgs vacuum expectation value to $\langle H \rangle = M_5 e^{-k L} \sim {\cal O} (1 \ \text{TeV})$. The fermionic content includes three copies of the left handed quark doublets, $\widehat{Q}^{i}, i= 1,2,3$, and three copies of right handed singlets, $\widehat{q}^{i} = \widehat{u}^{i},\widehat{d}^{i}$. Since the Clifford algebra in 5-dimension, given by five $4\times 4$ Gamma matrices ($\Gamma^M $), is non-reducible, the fermion representations in this geometry have 4 complex degrees of freedom. Hence, on breaking the 5-dimensional Lorentz group down to 4-dimensions via compactification these fermions become vector like under the Weyl representation of the 4-dimensional Clifford algebra. The orbifolding ensures that the unwanted set of chiralities are projected out from the lowest modes.

The five dimensional fermionic action for doublet($\widehat{Q}$) and singlet($\widehat{q}$) quarks with bulk mass terms is given as, 
\begin{eqnarray}\label{fermionAction}
 S_{\text{fermion}}& = &S_{\text{kin}} + S_{\text{yuk}}  \nonumber\\
S_{\text{kin}} &=& \int d^5 x \sqrt{-g}\, \left[\bar {\widehat{Q}} \left( \Gamma^M D_M + m_Q \right) \widehat{Q} + \sum_{q=u,d} \bar {\widehat{q}}  \left( \Gamma^M D_M
+ m_{q} \right) \widehat{q} \right] \\
S_{\text{Yuk}}&=& \int d^5 x \sqrt{-g}~ \bigg(\left(\tilde{Y}_{u}^{(5)}\right)_{ij}\bar {\widehat{Q}}_i \widehat{u}_j  + \left(\tilde{Y}_{d}^{(5)}\right)_{ij}\bar {\widehat{Q}}_i \widehat{d}_j\bigg)
H(x^\mu)\delta(y- L) + h.c. ,
\end{eqnarray}
where $m_Q$ and $m_q$ are the bulk masses for the doublet and singlet quark fields respectively; $i,j$ are  generational indices and $\left(\tilde{Y}_{u,d}^{(5)}\right)_{ij}$ are the five dimensional Yukawa matrices. In the above equation, $D_M$ represent the covariant derivative in 5-dimensions. The following boundary conditions at the orbifold fixed points $(y=0,y=L)$, ensure that only the correct chiral projections for the lowest modes of the respective fields survive at the boundaries.
\begin{equation}
\widehat{Q}_{l}(++),~\widehat{Q}_{r}(--),~\widehat{q}_{l}(--),~\widehat{q}_{r}(++) \ ,
\end{equation}
Here $l,r$ stand for left and right chiral fields under the 4-dimensional chiral projection operator and $+(-)$ stands for the Neumann (Dirichlet) boundary conditions.

The Kaluza-Klein decomposition of a generic fermion field (Q) is given by 
\begin{equation}
    Q(x,y)_{l,r} = \sum_{n =0}^\infty \frac{1}{\sqrt{L}} Q^{(n)}_{l,r}(x) f^{(n)}_{l,r}(y,c),
\end{equation}
where $Q^{(n)}_{l,r}(x)$ stands for the corresponding four dimensional chiral KK modes and  $f_{l,r}(y)$ are profiles of these modes in the bulk set to satisfy the ortho-normality conditions
\begin{equation}
\int_0^{L} dy \, e^{-3ky} \,f_{l,r}^{(n)}f_{l,r}^{(m)} = \delta_{n,m}.
\label{fermionnorm}
\end{equation}
The normalized zero mode profile for doublets and singlets with bulk mass parameter $c_{Q_{i}} = m_{Q_{i}}/k$ and $c_{q_{i}}=-m_{q_{i}}/k$ respectively is computed to be,
\begin{eqnarray}
f^{(0)}_{l}(y,c_{Q_{i}}) = \sqrt{k}f^0(c_{Q_{i}})\ e^{ky(2- c_{Q_{i}})}\ e^{(c_{Q_{i}}- 0.5)k L}\\
f^{(0)}_{r}(y,c_{q_{i}}) = \sqrt{k}f^0(c_{q_{i}})\ e^{ky(2- c_{q_{i}})}\ e^{(c_{q_{i}}- 0.5)k L}
\end{eqnarray}
where,
\begin{equation}
    f^0(c)= \sqrt{\frac{(1-2 c)}{1-e^{-(1-2c)k L}}} \ ,
    \label{fermionzeroprofile}
\end{equation}
In this paper we are not interested in the higher KK-modes of fermions, as their contribution will always be suppressed by $\sim {\cal O}(\langle H \rangle/M_{KK})$, where $M_{KK}$ is the compactification scale.
Inserting the zero mode profile into the Yukawa action given in Eq.(\ref{fermionAction}), we obtain the effective 4D Yukawa coupling relevant for the SM fermion masses and mixings as,
\begin{eqnarray}\label{Effective4DYukawa}
Y_{d_{ij}} =  f^{0}(c_{Qi})\left(Y_{d}^{(5)}\right)_{ij} f^{0}(c_{dj})\nonumber\\
Y_{u_{ij}} = f^{0}(c_{Qi}) \left(Y_{u}^{(5)}\right)_{ij} f^{0}(c_{uj})
\end{eqnarray}
where $\ord{(1)}$ Yukawa $(Y_{u,d}^{(5)})$ parameters entering the mass matrices are defined as:
\begin{equation}
Y_{u,d}^{(5)} \equiv k \tilde{Y}_{u,d}^{(5)}
\end{equation}
The transformation from the quark flavour eigenbasis $\widehat{u}_{l,r}, \widehat{d}_{l,r}$ to the mass eigenbasis  $u_{l,r}, d_{l,r}$ will then be given by performing rotation of unitary mixing matrices $U_{l,r}$ and $D_{l,r}$ as
\begin{eqnarray}
{u}_{l,r}=U_{l,r}^{\dagger}\widehat{u}_{l,r},~~~
{d}_{l,r}=D_{l,r}^{\dagger}\widehat{d}_{l,r}.
\label{massbasis}
\end{eqnarray}
With this the CKM matrix is given by
\begin{eqnarray}
V_{CKM}=U^{\dagger}_{l}D_{l}
\end{eqnarray}
It should be noted that there is no reason for bulk fermionic masses to be diagonal or real. Since we are not assuming Minimal Flavour Violation(MFV), the Unitary matrices that diagonalise these bulk mass terms do not diagonalise the five dimensional Yukawa Lagrangian. 

\paragraph{Gauge couplings}
To derive the bulk wave profile and coupling of the gauge boson with fermion bilinear it suffices to describe a $U(1)$ gauge group in the bulk of the AdS. The generalisation to non-abelian gauge fields is straight forward. The five dimensional action for such a gauge field is given as
\begin{eqnarray}
\mathcal{S}=- \frac{1}{4g_{5}^{2}} \int d^{5}x\sqrt{-g}\left(g^{CM}g^{DN}F_{CD}F_{MN} \right) \ ,
\label{Gluon}
\end{eqnarray}
where the field strength tensor $F_{MN}=\partial_{M}A_{N}-\partial_{N}A_{M}$ and  $g_{5}^{-2}$ is the 5D gauge coupling.

In the unitary gauge, the vector field can be Fourier expanded as 
\begin{equation}
    A_{\mu}(x,y)= \sum_{n}f^{(n)}_{A}(y)A_{\mu}^{(n)}(x),
\end{equation}
where $A_{\mu}^{(n)}(x)$ are the 4D gauge field KK modes and $f^{(n)}_{A}(y)$ are profiles of these modes in the bulk. The equations of motion are given as
by 

\begin{eqnarray}
-\partial_{5}\left(e^{-2ky}\partial_{5}f^{(n)}_{A}\right)=m^{2}_{n}f^{(n)}_{A} \ .
\label{gluoneq}
\end{eqnarray}
On integrating the above equation and demanding the zero KK mode be non vanishing, we arrive at the boundary condition $(\delta A^{\mu}\partial_{y}A_{\mu})\big|_{0,L}=0$. For canonical kinetic terms, the orthonormality condition is 
\begin{equation}
    \int_{0}^{L}dyf^{(n)}_{A}f^{(m)}_{A}=\delta_{nm}.
\end{equation}
Solving Eq.(\ref{gluoneq}) for $m_n=0$, we find that the zero-mode profile of the gauge boson is flat and is given as 
\begin{eqnarray}
f^{(0)}_{A}(y)=\frac{1}{\sqrt{L}}
\end{eqnarray}
For the higher KK modes ( $m_n \neq 0$) the solution is given in terms of the Bessel $J$ and $Y$ functions and is of the form:
\begin{eqnarray}
f^{(n)}_{A}(y)=N^{(n)}_{A}e^{ky}\left[J_{1}\left( \frac{m_{n}}{k e^{-ky}}\right) +b^{(n)}_{A}Y_{1}\left( \frac{m_{n}}{k e^{-ky}} \right)\right]
\label{gaugekkprofile}
\end{eqnarray}
$N_A^{(n)}$ are the the normalisation constants. The coefficients $b^{n}_A$ are determined at the boundaries as 
\begin{eqnarray}
b_{A}^{(n)}\big{|}_{y=0} = - \frac{J_{0}(\frac{m_{n}}{k})}{Y_{0}(\frac{m_{n}}{k}))}, ~~~~~
b_{A}^{(n)}\big{|}_{y=L} = - \frac{J_{0}(\frac{m_{n}}{k}e^{kL})}{Y_{0}(\frac{m_{n}}{k}e^{kL})}.
\end{eqnarray}
Equating $b_{A}^{(n)}\big{|}_{y=0}=b_{A}^{(n)}\big{|}_{y=L}$ we get the spectrum $m_{n}=x_{n}ke^{-kL}$ where $x_n$ are the roots of this equation.

\begin{figure}[h]
	\centering
	\includegraphics[height=6.5 cm,width=8 cm]{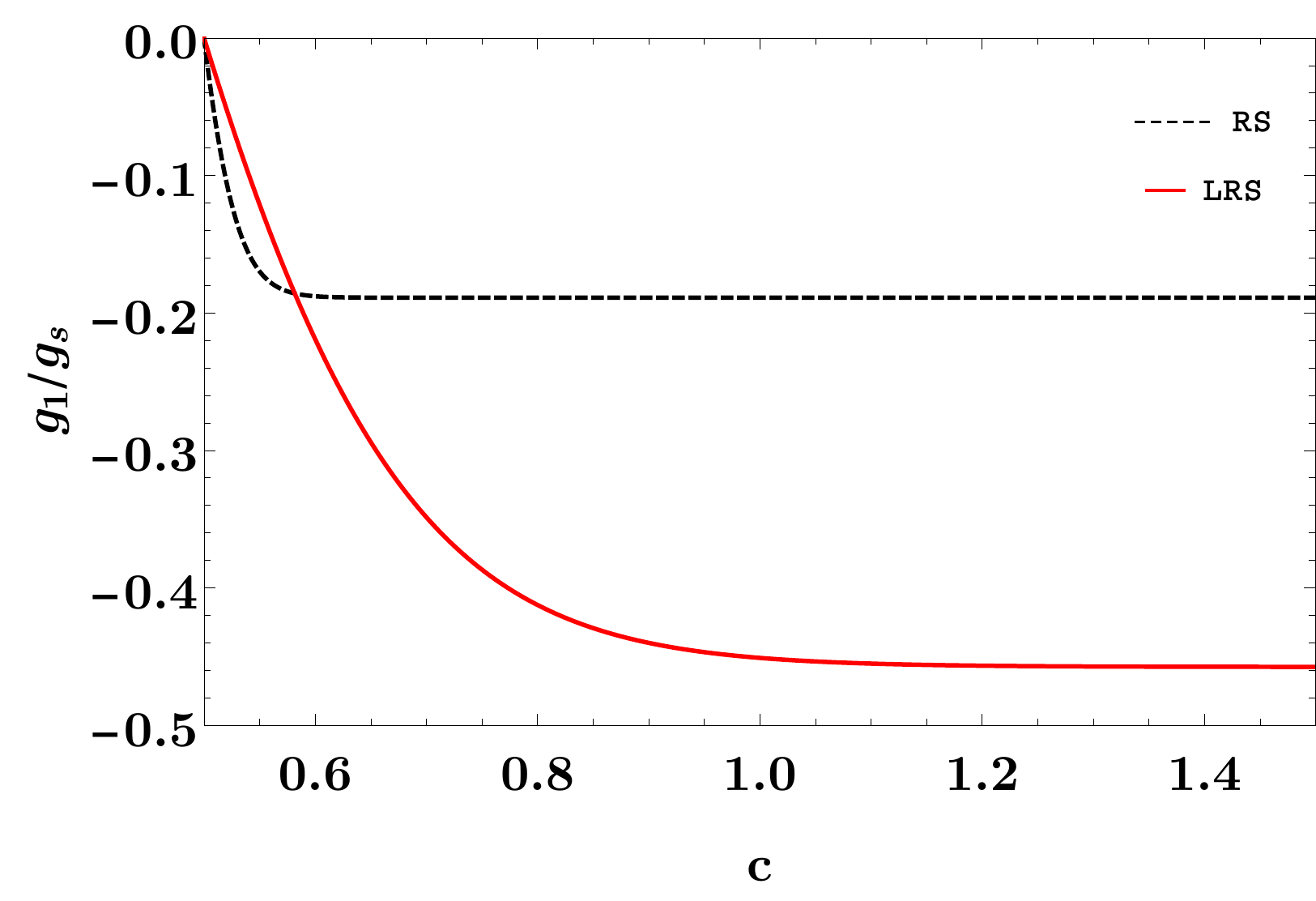}
	\caption{\sf A comparison of coupling of first KK gluon with light fermions as a function of the $c$ values in RS (blue/dotted) and Little RS (red/solid) }
	\label{RS_LRS_coupling}
\end{figure}

The coupling of the zero mode fermions with the gauge KK modes is set by the overlap integral
\cite{Gherghetta:2000qt}
\begin{eqnarray}\label{GaugeFOFO}
g_{l,r}^{(n)}(c_{Q,q})&=&g_5 \int_0^{L} dy \,e^{-3 k y} f_A^{(n)}(y) f_{l,r}^{(0)}(y,c_{Q,q}) f_{l,r}^{(0)}(y,c_{Q,q})
\end{eqnarray}
where $f_{l,r}^{(0)}$ and $f_A^{(n)}$ are given by Eq.(\ref{fermionzeroprofile}) and Eq.(\ref{gaugekkprofile}) respectively. Note that this function is quite sensitive to the UV scale 
and changing the scale could significantly modify the low energy phenomenology. Fig.(\ref{RS_LRS_coupling}) shows the difference between the coupling of KK-1 gluon in RS and Little RS and it could be seen that the coupling is enhanced by a factor 3 in the Little RS in comparison to RS for the fermion with the same '$c$' value. This could be understood from the fact that, since the fundamental scale is much smaller, the UV localized fermions have bigger overlap with the composite gauge boson states\footnote{For the IR localised fermions, one would expect the situation to be opposite, a reduction in the coupling.}. This has important implications for LHC phenomenology and $\epsilon_K$ as will be demonstrated shortly. 

A further important point regarding the coupling in Eq.(\ref{GaugeFOFO}) is that it is non-universal in generational space. This is evident when we explicitly write these couplings down as: 
\begin{eqnarray}
g^{(n)}_{l}(c_{Qi})\bar{\widehat{Q}}_{l}^{i(0)}\gamma_{\mu}G^{\mu(n)}{\widehat{Q}}_{l}^{i(0)}
+g^{(n)}_{r}(c_{ui})\bar{\widehat{u}}_{r}^{i(0)}\gamma_{\mu}G^{\mu(n)}{\widehat{u}}_{r}^{i(0)} + g^{(n)}_{r}(c_{di})\bar{\widehat{d}}_{r}^{i(0)}\gamma_{\mu}G^{\mu(n)}{\widehat{d}}_{r}^{i(0)}
\end{eqnarray} 
After electroweak symmetry breaking, unitary transformations ($D_{l,r}$, $U_{l,r}$) of Eq.(\ref{massbasis}) 
are used to go to mass eigen basis.This non-universality in couplings leads to flavour violation in these models. 

\paragraph{Phenomenological constraints on Little RS}
The original RS model is envisaged to solve the Higgs mass hierarchy problem. This soon led to phenomenological issues. Localising the entire SM to 3-brane located at IR predicted large flavour violating and proton decay currents. On allowing SM fields, except for the Higgs, to propagate into the bulk solved such large contributions. As an added benefit, doing so also relaxed the large S-parameter contribution to the Electro-weak observables~\cite{Agashe:2003zs,Carena:2003fx,Casagrande:2008hr,Iyer:2015ywa}. On the other hand, the T-parameter contribution remained large. And for a 125 GeV Higgs, the lower limit on mass of the lightest KK excitation turns out to be $\sim 13.6 \ \text{TeV} $~\cite{Iyer:2015ywa}. Thus, making this model irrelevant for the current LHC searches. Assuming a truncated space, Little RS ($k L \sim 7$) brings down this correction and lowers the limit of the compactification scale to $ \sim 4 \ \text{TeV}$~\cite{Davoudiasl:2008hx}.

Another important expectation from Little RS model comes from its accessibility at the colliders. At LHC, the production of the lightest KK gluon and the $Z$ boson in s-channel is via the annihilation of light quarks, which are typically UV localized. Fig.(\ref{RS_LRS_coupling}) clearly shows that the coupling of first KK gluon with light quark bilenears (bulk mass parameter $c \gtrsim 1.0$) are larger in Little RS in comparison with RS. This leads to enhanced production cross-section for this process. Recent searches were carried out at ATLAS collaboration~\cite{Aaboud_2018} which looks for a heavy particles that decay into top-quark pair at 13 TeV LHC with an integrated luminosity of 36.1$fb^{-1}$. And the data rules out the masses smaller than $\sim 3.6 $ TeV at 95\% CL for the KK gluon with a branching fraction of $30\%$. In this analysis, the strong coupling of light fermions to the KK gluon modes were set to $-0.2 g_s$, where $g_s$ is the strong coupling of the SM gluon. 
They considered the process $pp \to t\bar{t}$ with top decaying in to $t\to b W$ with events selected requiring single charged isolated lepton, jets, missing transverse energy (or $\not{p_T}$). SM background processes were reduced by requiring the jets identified as likely to contain b-hadrons. We use this result to constrain our model. 

\begin{figure}[h]
	\centering
	\includegraphics[height=14.0 cm,width=10.cm]{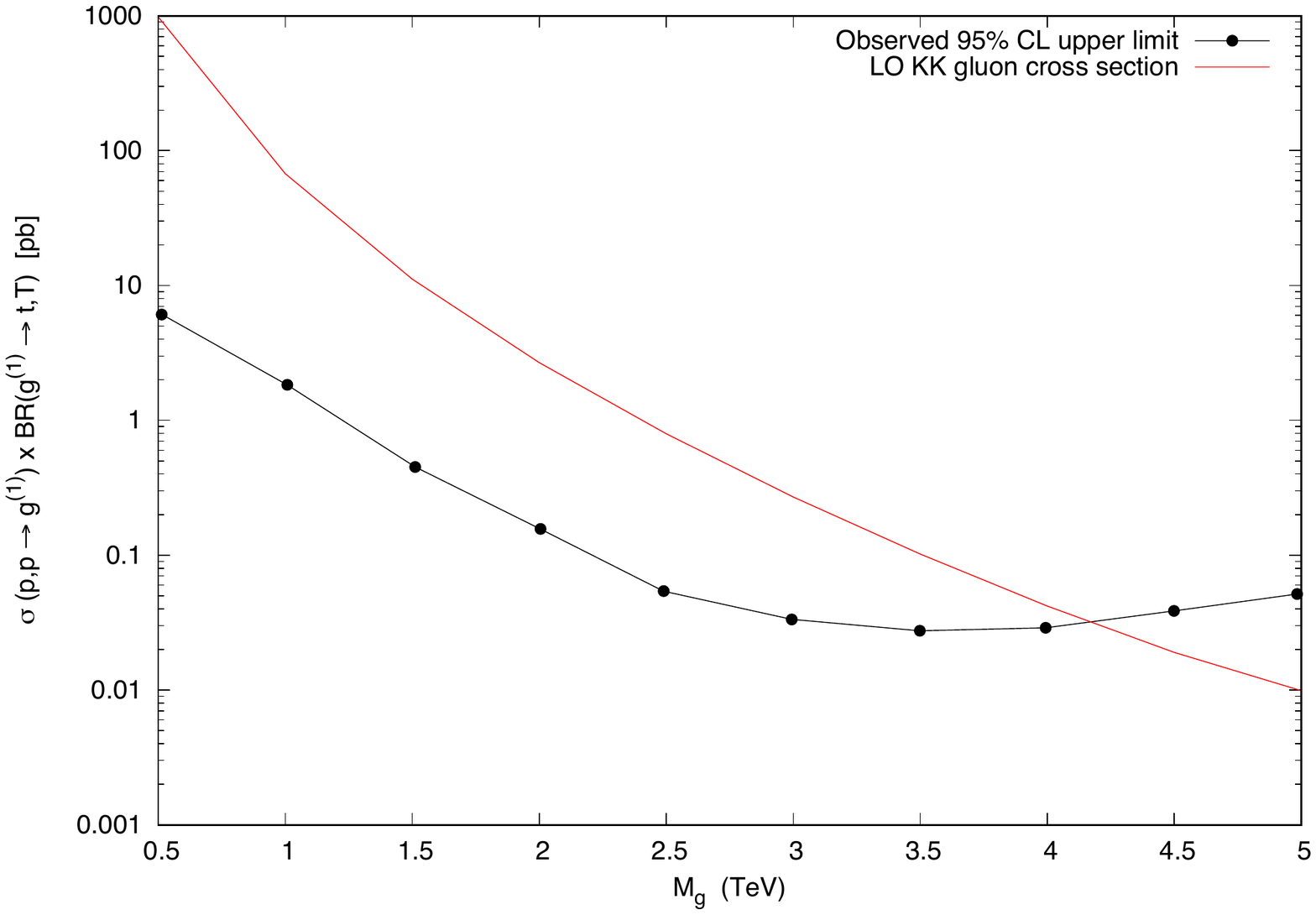}
	\caption{\sf The observed cross-section 95\% CL upper limit \cite{Aaboud_2018} on the $g_{1}$ signal and the theoretical prediction for the production cross-section times branching ratio of $g_1 \rightarrow t\bar{t}$ at corresponding masses}
	\label{LRSLHC}
\end{figure}

While these limits hold for the RS model, a straight forward interpolation to Little RS is only approximately valid. Still, a direct comparison to the limit given by ATLAS will give a conservative limit on mass of the lightest KK gluon in our model. For the analysis, we will use couplings and branching fractions demanded by the bulk mass parameters of our model given in Table \ref{BechPoint}. These parameters were chosen such that they would satisfy the CKM matrix and will be useful for our further analysis of $\epsilon_K$. 
 To compute the cross section of the process $q \ q \rightarrow g^{(1)} \rightarrow t \ \bar{t} $, we used CalcHEP 3.7.5 \cite{calchep} with NNPDF2.3 with QED corrections. Since we did not mention a specific K-factor, CalcHEP uses a version of this function which always returns 1. With these considerations we find that the mass of the first KK partner of gluon to be $\gtrsim 4.2 \ \text{TeV}$ (as shown in Fig.(\ref{LRSLHC}). And we will use this limit for rest of our analysis. 

\section{$\epsilon_K$ in Little RS}
\label{sec:littlers}
Other than the hadron collider and LEP limits mentioned in the previous section, observables from  the neutral meson mixing, especially the  CP-Violation observable ($\epsilon_K$) of $K^0 - \bar{K}^0$ system, has been shown to constraint the UV scale of RS \cite{Santiago:2008vq, Csaki:2008zd, Bauer:2009cf, Agashe:2008uz, Blanke:2008zb,Ahmed:2019zxm} and Little RS \cite{Bauer:2009cf, Bauer:2008xb} models significantly. As shown in Fig.(\ref{RS_LRS_coupling}), the light fermions couple to the first KK mode of gluons  more strongly in the case of Little RS compared to RS. Thus understanding this constraint in the Little RS setup is imperative. In this section, we recall the important results of $\Delta F =2$ process in warped geometry.
\begin{figure}[h]
	\centering
	\includegraphics[height=6.5 cm,width=10cm]{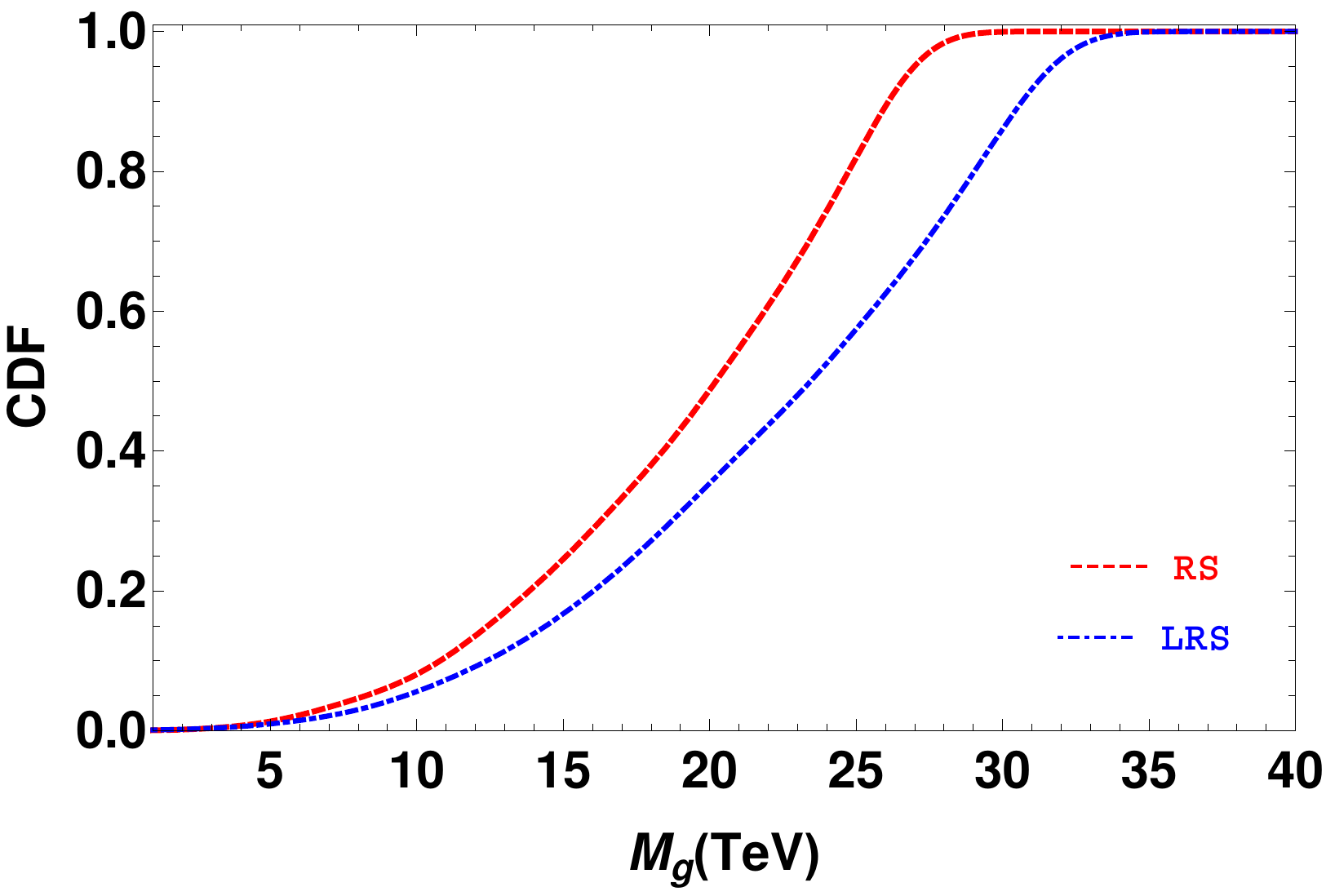}
	\caption{\sf CDF for $M_{g}$ from  $\epsilon_{K}$ (analysis performed from the dominant $|ImC_{4}^{sd}|$) for RS and LRS model. Dashed lines correspond to the RS and dash-dot to LRS.}\label{RS_LRS_Epsi}
\end{figure} 
Note that from Eq.(\ref{GaugeFOFO}) the coupling of the bulk gauge boson KK states to fermion bilinears are dependent on the bulk mass parameters of the fermions. This renders the couplings family non-universal in the gauge basis. On rotating to the mass basis, defined by $D_{l,r}$ and $U_{l,r}$ given in the Appendix, the couplings become flavour non-diagonal. In principle all the gauge KK states would contribute to this process, but the dominant contribution comes from the tree level exchange of the lightest gluon KK state. On integrating out the new physics with mass $M_{g}$, the effective Hamiltonian for $\Delta F=2$ becomes~\cite{Blanke:2008zb},

\begin{eqnarray}
H_{eff} =\frac{1}{M_{g}^{2}}\bigg[g_{l}^{ij}g_{l}^{km}(\bar{d}^{i\alpha}_{l}
T^{a}_{\alpha \beta}
\gamma_{\mu}d^{j\beta}_{l})(\bar{d}^{k\delta}_{l}
T^{a}_{\delta \rho}
\gamma_{\mu}d^{m\rho}_{l}) + g_{r}^{ij}g_{l}^{km}(\bar{d}^{i\alpha}_{r}
T^{a}_{\alpha \beta}
\gamma_{\mu}d^{j\beta}_{r})(\bar{d}^{k\delta}_{l}
T^{a}_{\delta \rho}
\gamma_{\mu}d^{m\rho}_{l}) + (l\leftrightarrow r)\bigg] \ ,
\label{effHamiltonian}
\end{eqnarray} 
where $T^{a}$ are the generators of the QCD gauge group, Latin indices $i,j,k,l,m$ denote the fermion generation index and Greek indices represent the colour indices. Couplings $\tilde{g}_{l,r}$ are defined as:
\begin{eqnarray}
\tilde{g}_{l} = D_{l}^{\dagger}g_{l}^{(1)}(c_{Q})D_{l},~~~~\tilde{g}_{r} = D_{r}^{\dagger}g_{r}^{(1)}(c_{d})D_{r} \ ,
\end{eqnarray}
where $g_{l}^{(1)}(c_{Q})$ and $g_{r}^{(1)}(c_{d})$ are the gluon couplings defined in Eq.(\ref{GaugeFOFO}). Using the unitarity of $D_{l,r}$, the above expression could be further expanded to understand the $\Delta F=2$ transitions in first and second generations. For this process, we only require the off diagonal elements which represent the flavour violating couplings and are given as,
\begin{eqnarray}
\label{Expressionglgr}
\tilde{g}_{l}^{12}&=&D_{l}^{(21)*}D_{l}^{(22)}\bigg(g_{l}^{(1)}(c_{Q2})-g_{l}^{(1)}(c_{Q1})\bigg)+D_{l}^{(31)*}D_{l}^{(32)}\bigg(g_{l}^{(1)}(c_{Q3})-g_{l}^{(1)}(c_{Q1})\bigg)\nonumber\\
\tilde{g}_{r}^{12}&=&D_{r}^{(21)*}D_{r}^{(22)}\bigg(g_{r}^{(1)}(c_{d2})-g_{r}^{(1)}(c_{d1})\bigg)+D_{r}^{(31)*}D_{r}^{(32)}\bigg(g_{r}^{(1)}(c_{d3})-g_{r}^{(1)}(c_{d1})\bigg)
\end{eqnarray}
Using Eq.(\ref{Expressionglgr}), the Fierz identity and the definitions of quadratic Casimir, the effective Hamiltonian \ref{effHamiltonian} could be simplified to
\begin{eqnarray}
\label{heffkaon}
H_{eff}&=&\frac{1}{M_{g}^{2}}\bigg[\tilde{g}_{l}^{ij}\tilde{g}_{l}^{km}\frac{1}{2} \bigg( (\bar{d}^{i\alpha}_{l}\gamma_{\mu}d^{j\delta}_{L})(\bar{d}^{k\delta}_{l}
\gamma_{\mu}d^{m\alpha}_{l})-\frac{1}{N_{C}} (\bar{d}^{i\alpha}_{l}\gamma_{\mu}d^{j\alpha}_{l})(\bar{d}^{k\delta}_{l}
\gamma_{\mu}d^{m\delta}_{l})\bigg) \\
&-& \tilde{g}_{r}^{ij}\tilde{g}_{l}^{km}((\bar{d}^{i\alpha}_{r}d^{m\alpha}_{l})
(\bar{d}^{k\delta}_{l}d^{j\delta}_{r})+\frac{1}{N_{C}}\big( (\bar{d}^{i\alpha}_{r}d^{m\delta}_{l})
(\bar{d}^{k\delta}_{l}d^{j\alpha}_{r})) + (l\leftrightarrow r) \bigg]\nonumber
\end{eqnarray}
Adopting the usual parametrization of new physics effects in Kaon oscillation~\cite{Ciuchini:1998ix, Ciuchini:1997bw} we can write the model independent effective Hamiltonian as
\begin{eqnarray}
H^{\Delta S=2} = \sum_{a=1}^{5} C_{a}\mathcal{O}_{a}^{sd} + \sum_{a=1}^{3} \tilde{C}_{a}\tilde{\mathcal{O}}_{a}^{sd}. 
\label{heffciuchini}
\end{eqnarray}
In the above equation, the first set of operators $C_{a}$ contain left handed chiral states and the operators $\tilde{C}_{a}$ contain right chiral states. And the operators ${\cal O}$ are,
\begin{eqnarray}
\mathcal{O}_{1}^{sd}&=&(\bar{d}_{l}^{\alpha}\gamma_{\mu}s_{l}^{\alpha})(\bar{d}_{l}^{\beta}\gamma_{\mu}s_{l}^{\beta}),\\
\mathcal{O}_{4}^{sd}&=&(\bar{d}_{r}^{\alpha}s_{l}^{\alpha})(\bar{d}_{l}^{\beta}s_{r}^{\beta}),~~~
\mathcal{O}_{5}^{sd}=(\bar{d}_{r}^{\alpha}s_{l}^{\beta})(\bar{d}_{l}^{\beta}s_{r}^{\alpha}),\nonumber
\label{kops}
\end{eqnarray}
and $\tilde{\mathcal{O}}_{i}$ is given by exchanging $l\leftrightarrow r$ in $\mathcal{O}_{i}$.
Comparing the above expressions for the operators with the Effective Hamiltonian given in Eq.(\ref{heffkaon}), we could infer 
\begin{eqnarray}\label{coupling}
C_{1} &=& \frac{1}{M_g^{2}}\tilde{g}^{12}_{l}\tilde{g}^{12}_{l}\bigg[\frac{1}{2}(1-\frac{1}{N_{C}})\bigg],~~~
C_{4} = \frac{1}{M_g^{2}}\tilde{g}^{12}_{l}\tilde{g}^{12}_{r}\big[-1\big],~~~
C_{5} = \frac{1}{M_g^{2}}\tilde{g}^{12}_{l}\tilde{g}^{12}_{r}\big[\frac{1}{N_{C}}\big] \ .
\label{EqC1C4C5}
\end{eqnarray}
Since the CP-Violating $\epsilon_K$ parameter in terms of the Effective Hamiltonian Eq.(\ref{heffciuchini}) is given as,
\begin{equation}
    \epsilon_K \propto Im \langle K^0|H^{\Delta S=2}| \bar{K^0} \rangle \ ,
\end{equation}
only the imaginary parts of the Wilson coefficients contribute and the bound on  the scale $M_{g}$ is summarised in Table (\ref{tablewilson})

\begin{table}[ht]
\centering
\begin{tabular}{c|c}
Im(Wilson coeff.) & Bound ($\text{TeV}$) \\
\hline
  Im$(C_1)$& $1.5 \times 10^3$\\
   Im$(C_4)$  &$1.6 \times 10^4 $ \\
  Im$(C_5)$ & $1.4 \times 10^4$ \\
  \hline
\end{tabular}
\label{}
  \caption{Lower limit on  Wilson coefficients at scale $3$ TeV \cite{Csaki:2008zd}.}
  \label{tablewilson}
\end{table}
The model independent bound from $\epsilon_{K}$ is strongest on the Wilson coefficient $C_4$ due to  (as compared to $C_1$ ) 
chiral enhancement of the hadronic matrix element and 
the  RG running from the new physics scale
to the hadronic scale~\cite{ Ciuchini:1998ix,Bagger_1997,Buras_2000}. 
With the above formalism in place, comparison of the constraint coming from $\epsilon_K$ in RS and Little RS model would be rendered simple since the difference between these models arise through the off diagonal couplings $\tilde{g}^{12}_{l,r}$ in the respective scenarios. For numerical study, we fit the bulk quark mass parameters to obtain the spectrum in $\overline{\text{MS}}$ scheme computed at $3 \ \text{TeV}$ while keeping the 5D Yuakwa anarchic ($0.1\le|Y_{{u,d}_{ij}}^{(5)}|\le3$). We refrain from discussing the procedure here, but the detailed methodology for the numerical analysis can be found in Appendix~\ref{FlavouParameters}. 

In Fig.(\ref{RS_LRS_Epsi}), we present our result showing the dependence of the cumulative distribution functions (CDF) for the number of states satisfying the $\epsilon_K$ observable ($|Im (C_{4}^{sd})|$) with varying KK gluon mass scale $M_{g}$ for both RS and Little RS. It can be seen that the average value of  $\epsilon_{K}$ becomes consistent with the measurement only for $M_{g} \ge 24$ TeV for RS and $M_{g} \ge 32$ TeV for Little RS.
From Eq.(\ref{GaugeFOFO}) and Eq.(\ref{coupling}) one could infer that the bounds on the operators $C_{1,4,5}$ crucially depends on the value of $g_{5}$ and the fermion bilinear overlap with the first KK gluon wavefunction. Reducing this overlap can be achieved by using Brane Localized Kinetic Terms.
This possibility has been mentioned earlier in Refs. \cite{Csaki:2008zd, Agashe:2008uz} within the context of RS models.
\label{phases}

\section{Brane Localized Gauge Kinetic Terms}
\label{sec:BLKTLRS}
To understand the impact of Brane Localized Kinetic Term (BLKT) on the Wilson coefficients given in Eq.(\ref{EqC1C4C5}), lets start by considering the U(1) gauge field in the warped background. This could easily be generalized to the case of non-Abelian fields, since we are only interested in operators in the fermion bilinears. The generalisation to the gauge 5-dimensional action given in Eq.(\ref{Gluon}), including the BLKT could be written as\cite{Carena:2002dz,Carena:2003fx}
\begin{eqnarray}
\mathcal{S}=- \frac{1}{4g_{5}^{2}} \int d^{5}x\sqrt{-g}\left(g^{AM}g^{BN}F_{AB}F_{MN} + [l_{IR} \delta(y-L) + l_{UV} \delta(y)]g^{\alpha \mu }g^{\beta \nu}F_{\alpha \beta}
F_{\mu \nu} \right) \ ,
\label{GluonAction}
\end{eqnarray}
where, $l_{IR}$ and $l_{UV}$ are the strengths of the localized kinetic terms at IR and UV branes respectively. Here, we have chosen the convention where both the 5-dimensional gauge field  $A_\mu(x,y)$ and  $g_{5}^{-2}$ have dimensions of mass. 

\begin{figure}[h]
	\centering
	\subfigure[]{
	\includegraphics[height=6.5 cm,width=8cm]{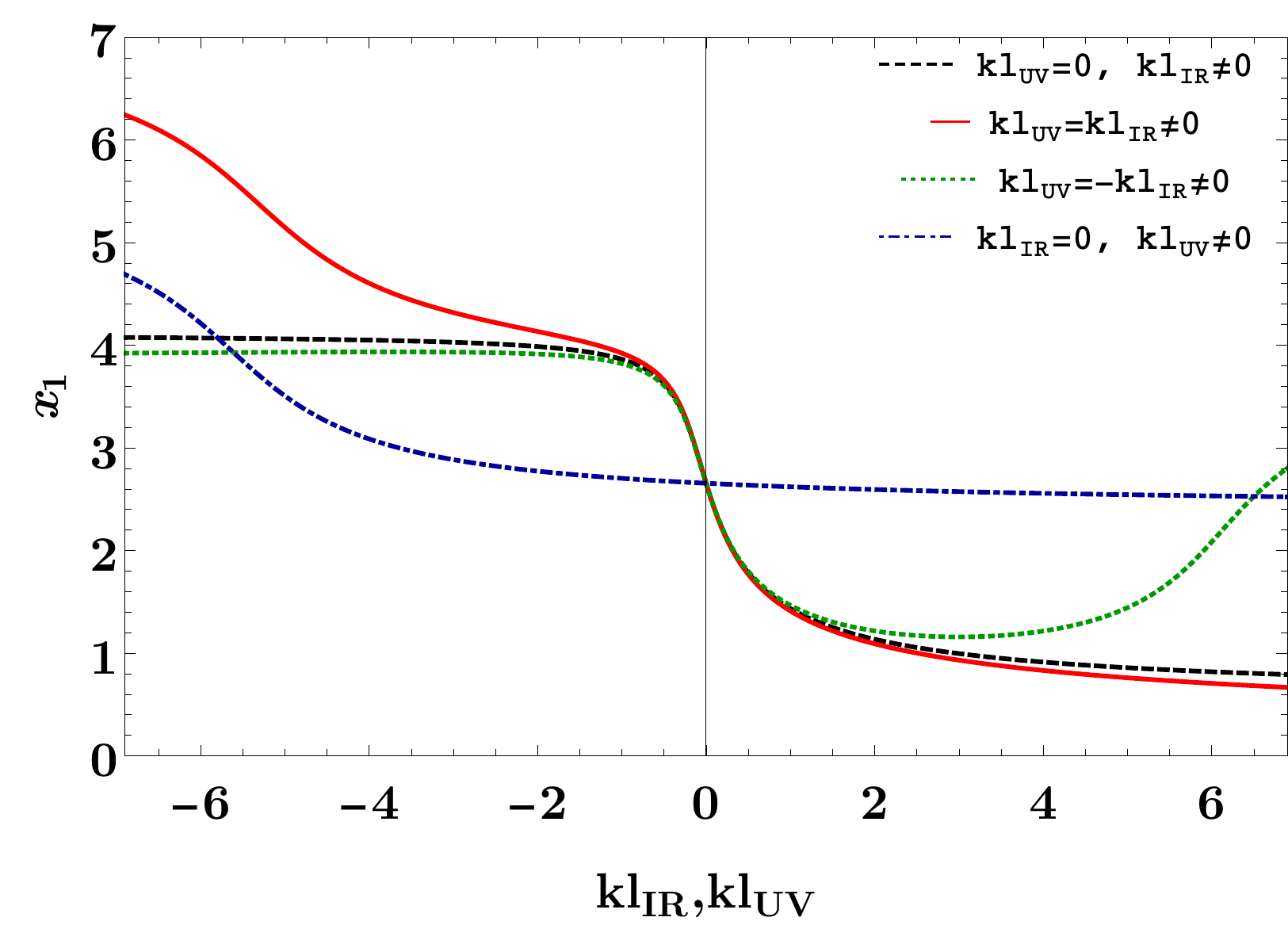}}
\label{EigenValu}
	\subfigure[]{
		\includegraphics[height=6.5cm,width=8cm]{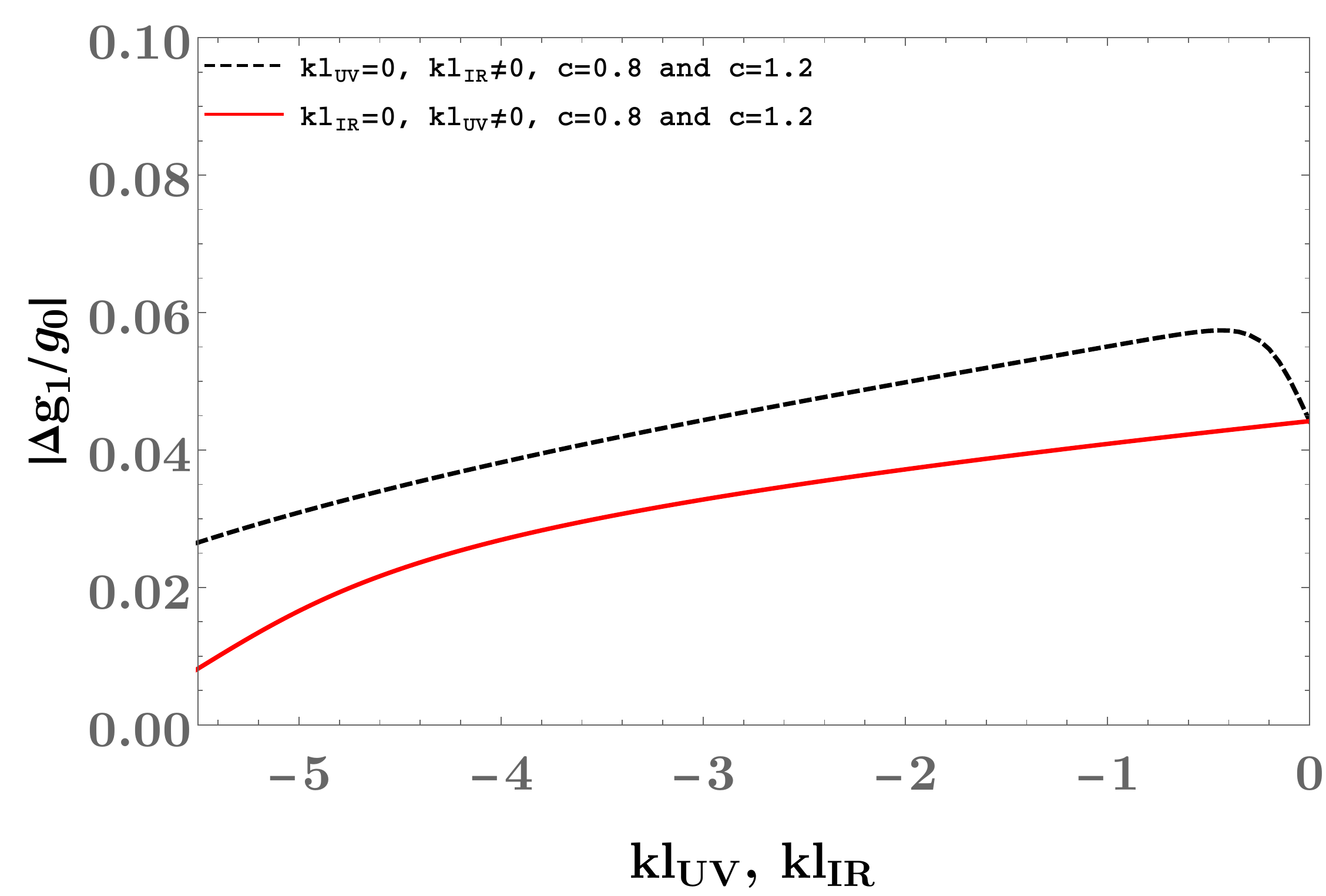}}
	\caption{\sf (a) The behaviour of the first root as function of $l_{IR}$ and $l_{UV}$. We have considered four cases:(i) $kl_{UV}=0$ (black dashed),(ii) $kl_{IR}=0$ (blue dot-dashed),(iii) $kl_{IR}=kl_{UV}$ (red solid) and (iv) $kl_{IR}=-kl_{UV}$ (green dotted) (b) The difference in KK-1 gluon gauge couplings, $|\Delta g_1/g_0|$,with fermions with bulk mass parameters $c=0.8$ and $c=1.2$ as a function of $kl_{UV}$ (assuming $kl_{IR} = 0$ (red solid)) and $kl_{IR}$ (assuming $kl_{UV}= 0$ (black dashed)).}\label{G1G0Couplings}
\end{figure}
In the presence of BLKT, the equation of motion derived from Eq.(\ref{GluonAction}) becomes,
\begin{eqnarray}
-\partial_{5}(e^{-2ky}\partial_{5}f^{(n)}_{A})=(1+l_{IR}\delta(y-L)+l_{UV}\delta(y))m^{2}_{n}f^{(n)}_{A} \ .
\label{diffeq}
\end{eqnarray}
And the ortho-normality condition is,
\begin{eqnarray}
\int_{0}^{L}dy\left[1+l_{IR}\delta(y)+l_{UV}\delta(y-L)\right]f^{(n)}_{A}f^{(m)}_{A}=\delta_{nm}.
\end{eqnarray}
The solution to the Eq.(\ref{diffeq}) is given by
\begin{eqnarray}
f^{(n)}_{A}(y)=N^{(n)}_{A}e^{ky}\left[J_{1}\left( \frac{m_{n}}{k e^{-ky}}\right) +b^{(n)}_{A}Y_{1}\left( \frac{m_{n}}{k e^{-ky}} \right)\right]
\label{blktfa}
\end{eqnarray}
where ($N^{(n)}_{A}$) is the normalisation constant for the n$^{th}$ mode and zero mode wavefunction is given by,
\begin{eqnarray}
f^{(0)}_{A}(y)=\frac{1}{\sqrt{L+l_{IR}+l_{UV}}} \ .
\end{eqnarray} 
Integrating the equation of motion at the fixed points $y=0$ and $y=L$ yields the modified boundary conditions
\begin{eqnarray}
\partial_{y}f^{(n)}_{A}|_{0} &=&  - l_{UV}m_{n}^{2}f^{(n)}_{A}(0)\nonumber \ ,\\
\partial_{y}f^{(n)}_{A}|_{L} &=&  + e^{2kL}l_{IR}m_{n}^{2}f^{(n)}_{A}(L).
\label{EigenSolve}
\end{eqnarray}
Demanding that the solution Eq.(\ref{blktfa}) should satisfy the boundary conditions, the $b_{A}^{(n)}$ becomes,
\begin{eqnarray}
b_{A}^{(n)}|_{UV} &=& - \frac{J_{0}(\frac{m_{n}}{k})+m_{n}l_{UV}J_{1}(\frac{m_{n}}{k})}{Y_{0}(\frac{m_{n}}{k})+m_{n}l_{UV}Y_{1}(\frac{m_{n}}{k})} \ ,\\
b_{A}^{(n)}|_{IR} &=& - \frac{J_{0}(\frac{m_{n}}{k}e^{kL})-m_{n}l_{IR}e^{kL}J_{1}(\frac{m_{n}}{k}e^{kL})}{Y_{0}(\frac{m_{n}}{k}e^{kL})-m_{n}l_{IR}e^{kL}Y_{1}(\frac{m_{n}}{k}e^{kL})} \ ,
\end{eqnarray}
where $m_{n}=x_{n}ke^{-kL}$, and $x_{n}$ are the roots of the master equation obtained by imposing $b_{A}^{(n)}|_{UV}$=$b_{A}^{(n)}|_{IR}$. Now, we try to understand the implications of BLKTs on the KK-spectrum. The dependence of the first root, $x_{1}$, on BLKT is shown in the Fig.(\ref{G1G0Couplings}(a)) for the Little RS geometry with warp factor $k=10^{3}$ TeV. We have considered four cases of
BLKTs (i) $kl_{UV}$ = 0, $kl_{IR} ~ \neq 0$, (ii) $kl_{UV}$ = $kl_{UV} \neq 0$, (iii) $kl_{UV}$ = - $kl_{UV} \neq 0$ and (iv) $kl_{UV} \neq 0$, $kl_{IR} = 0$. As can be seen 
the lowest root itself can modify by roughly a factor between 2 and 3 for the case (ii), when both BLKTs are switched on, with the same sign\footnote{This in fact the largest variation one can expect in the spectrum, from the case without BLKT for which the root $x_1\sim 2.7$. }. 

\begin{figure}[h]
	\centering
	\subfigure[]{
		\includegraphics[height=5.5 cm,width=7.cm]{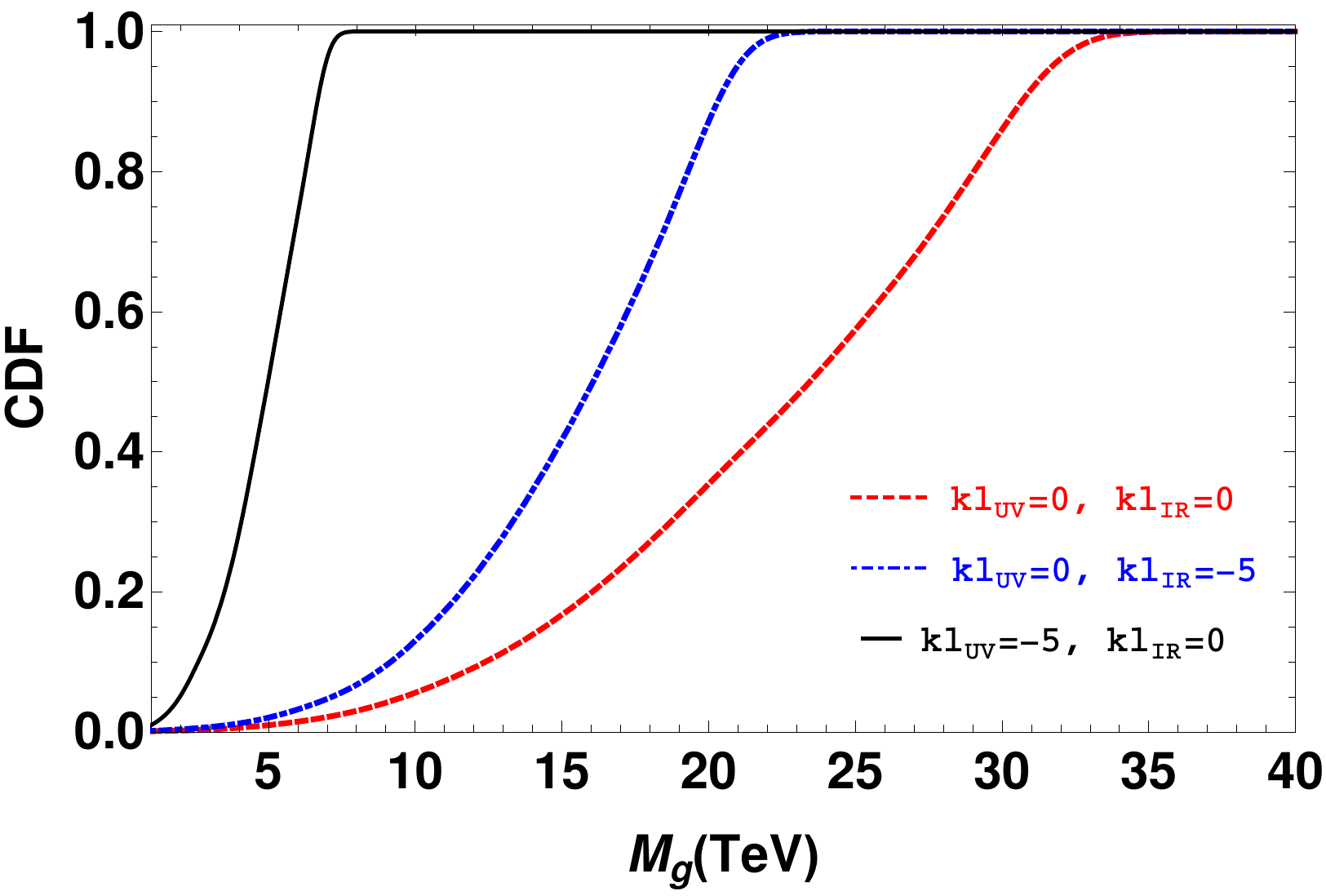}}
	\subfigure[]{
		\includegraphics[height=5.5 cm,width=7.cm]{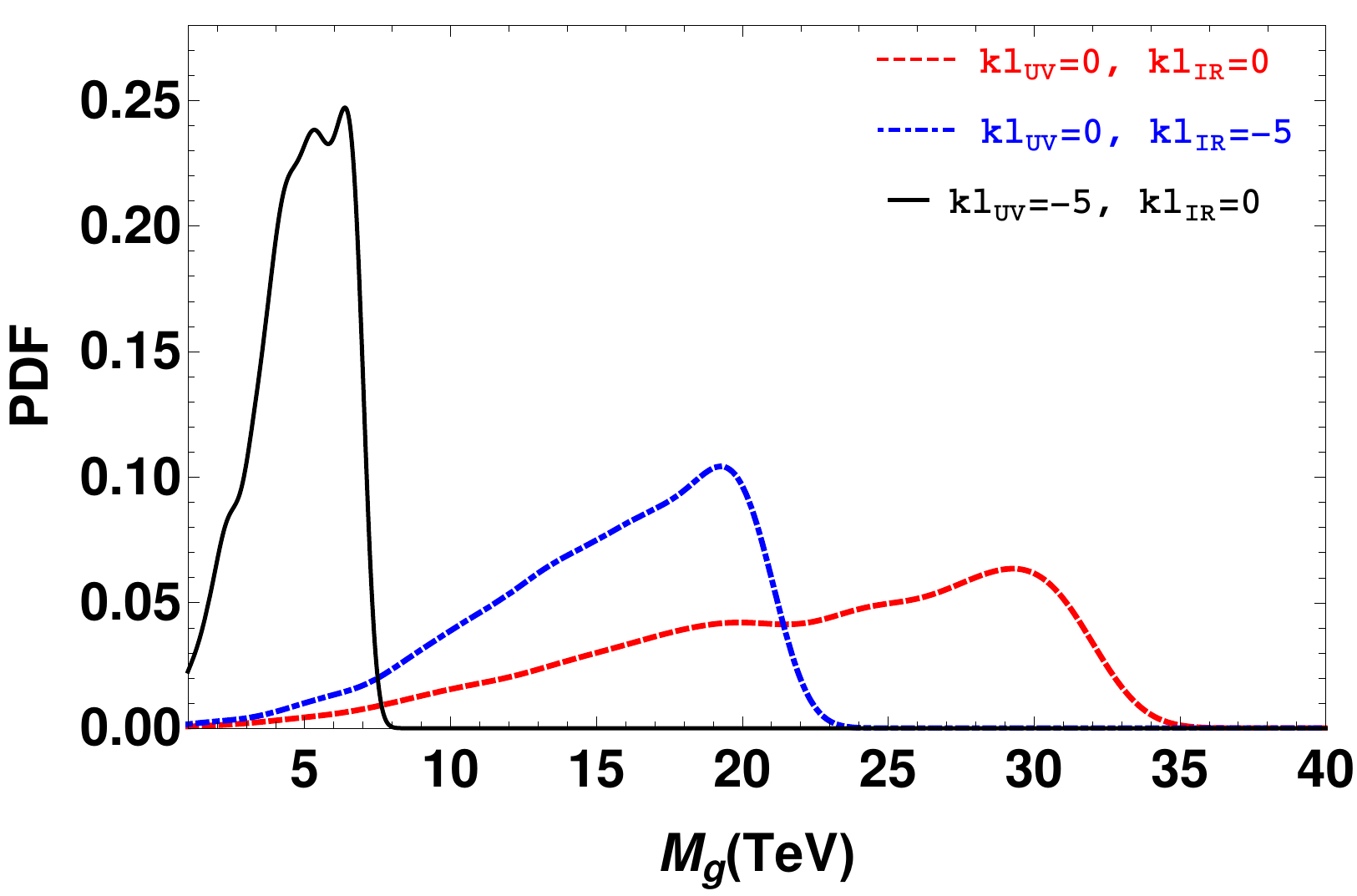}}
	\caption{\sf Cumulative distribution function and Probability distribution function satisfying the constraint on $Im(C_4)$ for three different scenarios: (a) $kl_{IR}=kl_{UV}=0$ (red dash) (b) $kl_{UV}=0,kl_{IR}=-5.0$ (blue dot-dash)(c) $kl_{IR}=0,kl_{UV}=-5$.(black solid)}\label{Plots1}
\end{figure}
A bigger change is however expected in the coupling of KK-gluon and the fermion bilinear since BLKT modifies the bulk gauge field wave profile as shown in Eq.(\ref{blktfa}) and hence the overlap of the gauge boson with the fermion wave profile. With that, the gauge coupling becomes,
\begin{eqnarray}
g_{l,r}^{(n)}(c_{Q,q})&=&g_5 \int_0^{L} dy \,e^{-3 k y} f_A^{(n)}(y) f_{l,r}^{(0)}(y,c_{Q,q}) f_{l,r}^{(0)}(y,c_{Q,q})\nonumber\\
&=&g_4\sqrt{L+l_{IR}+l_{UV}} \int_0^{L} dy \,e^{-3 k y} f_A^{(n)}(y) f_{l,r}^{(0)}(y,c_{Q,q}) f_{l,r}^{(0)}(y,c_{Q,q}) \ .
\label{coup_BLKT}
\end{eqnarray}
This modification could have significant impact on flavour observable. Following the discussion in Sec.\ref{sec:littlers}, we plot the relevant coupling for the $\epsilon_K$ process as a function of BLKT strength. To this extent we define 
\begin{equation}
\Big|\frac{\Delta g_1}{g_0}\Big| = \Big|\frac{g_{l,r}^{(1)}(c_2)-g_{l,r}^{(1)}(c_1)}{g_0}\Big| \ ,
\label{deltag1}
\end{equation}
where $g_{l,r}^{(1)}(c)$ is given in Eq.(\ref{coup_BLKT}). 
Fig.(\ref{G1G0Couplings}(b)) displays the difference in coupling strengths of first KK partner of the gluon, $\Big| \frac{\Delta g_1}{g_0}\Big|$, with fermions localized with the bulk mass parameters $c_{1}=1.2$ and $c_{2}=0.8$. In the figure we have explicitly shown the scenarios with $kl_{IR} = 0, kl_{UV}\neq0$ and vice versa. The values of $\Big| \frac{\Delta g_1}{g_0} \Big|$ for (i) $kl_{IR}=kl_{UV}=0$ (ii) $kl_{IR}= -5$,$kl_{UV}=0$ and (iii) $kl_{IR}=0$,$kl_{UV} = -5$ are given in Table \ref{numericalcoupling}. This clearly shows that UV BLKT is significantly better in reducing the couplings and considerably relaxes the bound in Little RS.
\begin{table}[h]
\centering
	\renewcommand{\arraystretch}{1.4}
	\begin{tabular}{|c|c|c|c|}
		    \hline
	    	    BLKT & $kl_{IR}=kl_{UV}=0$ & $kl_{IR}= -5$,$kl_{UV}=0$ & $kl_{IR}=0$,$kl_{UV} = -5$  \\
		\hline
		
		$\Big| \frac{\Delta g_1}{g_0} \Big|$ & 0.044 & 0.030 & 0.016  \\
		\hline
	\end{tabular}
	\caption{$\Big| \frac{\Delta g_1}{g_0} \Big|$ values for the three cases of BLKTs computed with $c_{1}=1.2$ and $c_{2}=0.8$.}
	\label{numericalcoupling}
\end{table}

For the full numerical analysis, we have used the fermion masses given in Table \ref{fermionmass} and scanned over the bulk mass parameter, 'c', with central values given in Table \ref{BechPoint}. This was done while imposing the anarchic condition on Yukawa, $0.1 \leq |Y^{(5)}_{u,d_{ij}}|\leq 3$. Fermion masses were fit  as detailed in the Appendix[\ref{FlavouParameters}]. The Cumulative distribution function and Probability distribution function satisfying the constraint on $Im(C_4)$ is presented in Fig.(\ref{Plots1}). The plots clearly show the drastic reduction on the constraints while imposing BLKTs. The result is even more spectacular for a UV BLKT. 
The bounds for BLKT$\sim 5$ are  summarised in Table \ref{Benchmark}. This clearly shows that the UV BLKT is significantly better in reducing the couplings and thus relaxes the bounds. 
It is clear that the BLKTs achieve this by bringing the couplings of the two light flavours closer to each other. Thus it is only natural that such a thing could also happen by imposing a flavour symmetry which we will explore in the next section.

\begin{table}[h]
\centering
	\renewcommand{\arraystretch}{1.4}
	\begin{tabular}{|c | c | c |c | c | c |c  |c |c|c|}
		\hline
		Parameter & $c_{Q1}$ & $c_{Q2}$ & $c_{Q3}$ & $c_{d1}$ & $c_{d2}$ & $c_{d3}$ & $c_{u1}$ & $c_{u2}$ & $c_{u3}$\\
		\hline
		c  & 1.21 & 1.13 & 0.36 & 1.29 & 1.15 & 1.02 & 1.65 & 0.59 & -0.80\\
		\hline
	\end{tabular}
	\caption{Central values 'c' parameter of the fermions used to obtain the fits in Table \ref{Benchmark}.}
	\label{BechPoint}
\end{table}
\begin{table}[h]
\centering
	\renewcommand{\arraystretch}{1.4}
	\begin{tabular}{|c|c|c|}
		    \hline
	    	    $kl_{IR}=kl_{UV}=0$ & $kl_{IR}= -5$,$kl_{UV}=0$ & $kl_{IR}=0$,$kl_{UV} = -5$  \\
		\hline
		
		30 TeV & 20 TeV  & 8 TeV \\
		\hline
	\end{tabular}
	\caption{Bounds on the KK-1 gluon masses for the different BLKTs considered. }
	\label{Benchmark}
\end{table}
\section{Minimal Flavour Protection}
Minimal Flavour Protection (MFP) \cite{Santiago:2008vq} was introduced to suppress the chiral enhanced New Physics contributions
to the $\Delta F =2$ Hamiltonian. These are typically contained in $C_4$ and $C_5$ Wilson coefficients, while the dominant Standard Model contributions are contained in $C_1$. In MFP, the singlet down sector is assumed to transform as triplet under a global $U(3)$ group with all other fields transforming as singlets. Hence the bulk mass term of fermions, taken to be diagonal in the flavour basis, become 
\begin{equation}
    c_{Q i} \bar{Q}_i Q_i + c_{u i} \bar{u}_i u_i + c_{d}\bar{d}_i d_i \ ,
\end{equation}
where $ Q_i$ and $u_i$, $d_i$ are the respective doublet and singlet quark fields in five dimensions. Note that the $U(3)$ symmetry makes sure that the bulk mass parameter $c_d$ is the same for all three generations of the singlet down sector. Hence unlike the scenario discussed in the previous section, the hierarchy in singlet down sector wave profiles, given in Eq.(\ref{eq:para1}), vanishes.
Due to which the couplings of n$^{th}$ KK-gluon with the right handed chiral zero mode bilinears, as given in Eq.(\ref{GaugeFOFO}), become
\begin{equation}
    \Big(g^{(n)}_r(c_d)\Big)_{i j} = g^{(n)}_d \delta_{i j} \ ,
\end{equation}
where $i, j$ denote the generation index and $g^{(n)}_d$ the corresponding coupling. And, from the second line of Eq.(\ref{Expressionglgr}), this leads to
\begin{equation}
    \tilde{g}_{r}^{12}= 0 \ .
\end{equation}
Hence this paradigm expects a conserved chiral symmetry with a consequence of significantly suppressing the contributions from $C_4$ and $C_5$. Its application to RS model is discussed in literature\cite{Santiago:2008vq}. 

\begin{figure}[h]
	\centering
	\subfigure[]{
		\includegraphics[height=5.5 cm,width=7.cm]{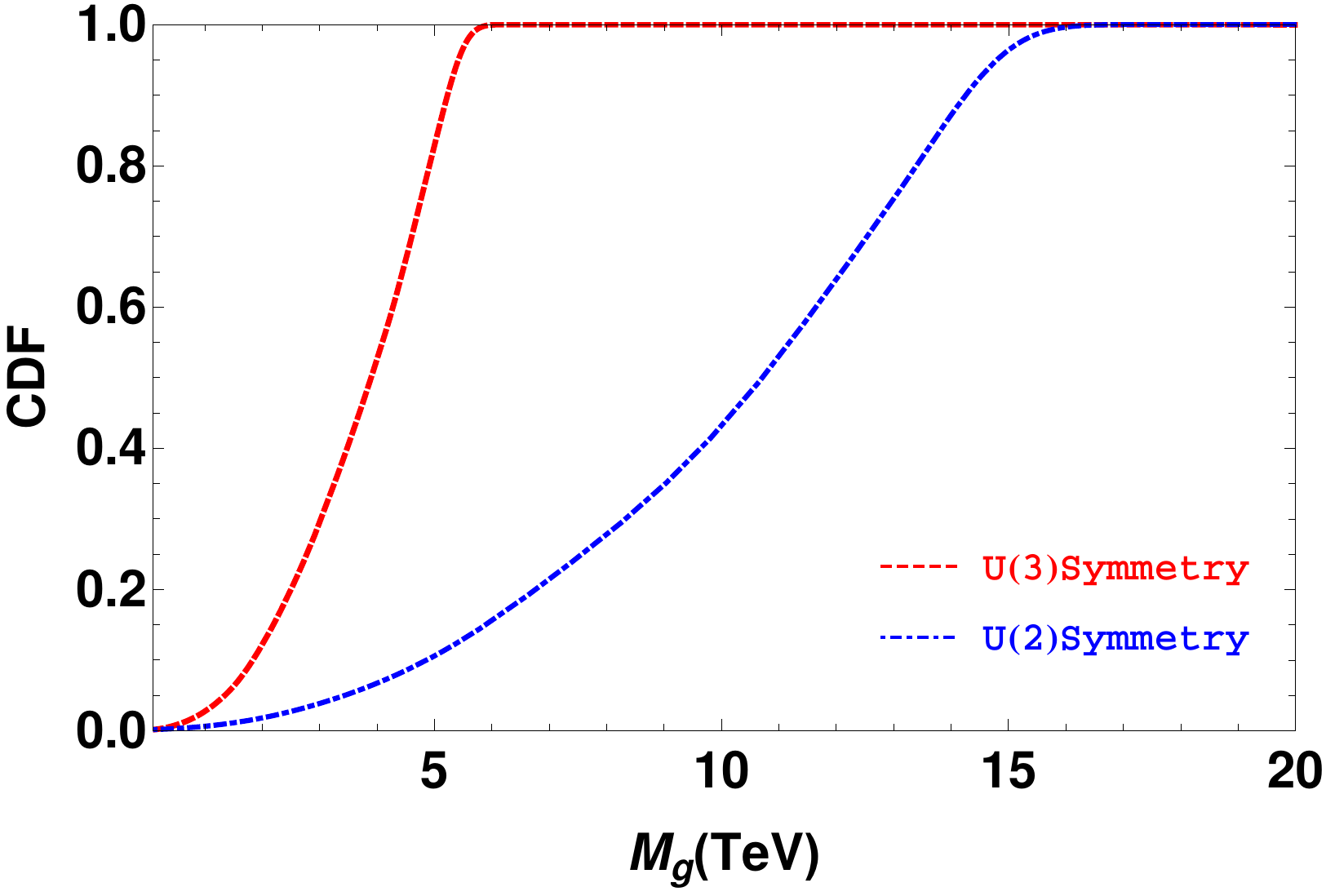}}
	\subfigure[]{
		\includegraphics[height=5.5 cm,width=7.0cm]{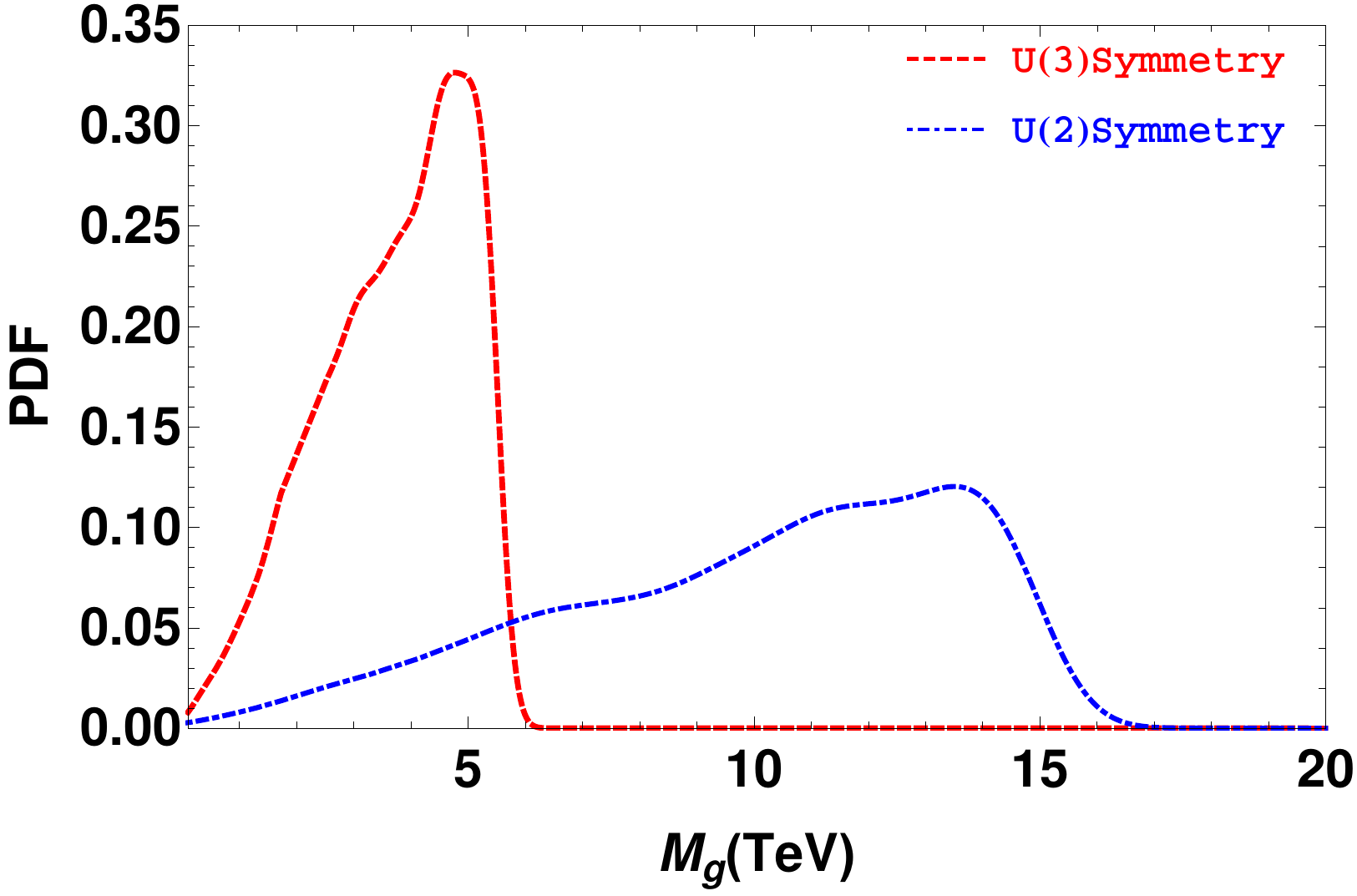}}
	\caption{
	 Cumulative distribution function (a) and Probability distribution function (b) regions satisfying the constraint on 
	 $Im(C_1)$ (for the cases of Unbroken U(3) flavour symmetry (red dashed)), and $Im(C_4)$ (for the cases of Unbroken U(2) flavor symmetry (blue dot-dash)) as a function of $M_g$.} \label{PlotsRSMFP}
\end{figure}

On the other hand, in Little RS, though the most important contribution comes again from the $C_4$ coefficient, the bound is more stringent than in RS as shown in Fig.(\ref{RS_LRS_Epsi}). And the lowest allowed mass of the first KK excited gluon is $M_g \gtrsim 35$ TeV. To solve this, we use the same horizontal $U(3)$ symmetry as defined above to prohibits such operators at first order. Though higher order operators are generated at dimension-8 arise due to Higgs insertion, but they can be neglected. And hence in all practicality $C_4$ operator does not play a role in the limit of exact $U(3)$ symmetry.

Here we consider two different scenarios, one with exact $U(3)$ horizontal symmetry and the one where this symmetry gets broken down to a $U(2)$ involving the first two generations. 
When $U(3)$ flavour symmetry is assumed to be exact, the bulk mass parameters for the right handed down sector becomes $c_{d1}=c_{d2}=c_{d3}$. We can observe from Eq.(\ref{Expressionglgr}), like in RS, the non diagonal coupling $\tilde{g}_{r}^{12}$ vanish, hence only $C_{1}$ operator contributes to the observable. We do the similar numerical analysis as in the previous section, but now imposing the flavor symmetry. 
The results are presented in  Fig.(\ref{PlotsRSMFP}) where CDF ( left) and PDF(right) are presented. As can be seen,  the bounds for LRS-MFP gets lowered to $ M_g \sim 4$ TeV. 
On the other hand, in the case where the full symmetry is broken down to $U(2)$, the non-diagonal coupling corresponding to flavour violation in s and d quark given in Eq.(\ref{Expressionglgr}) gets modified to
\begin{eqnarray}\label{Expressiongr}
\tilde{g}_{r}^{12}&=&D_{r}^{(31)*}D_{r}^{(32)}\bigg(g_{r}^{(1)}(c_{d3})-g_{r}^{(1)}(c_{d1})\bigg)
\end{eqnarray}
In this consideration, the bounds on first KK gluon mass becomes $M_g \gtrsim 18$ TeV.
Using the naive expression of $D_{r}$ from Appendix[\ref{FlavouParameters}] \bigg($|D_{r}|_{ij} \sim \frac{f^{(0)}(c_{di})}{f^{(0)}(c_{dj})}$\bigg), the $\tilde{g}_{r}^{12}$ is given as 
\begin{eqnarray}
\tilde{g}_{r}^{12} \sim \frac{f^{(0)}(c_{d1})}{f^{(0)}(c_{d3})}\frac{f^{(0)}(c_{d2})}{f^{(0)}(c_{d3})}\bigg(g_{r}^{(1)}(c_{d3})-g_{r}^{(1)}(c_{d1})\bigg)
\end{eqnarray}
where $f^{(0)}(c_{di})$ dictates the masses of quark in down sector  as shown in Eq.(\ref{Effective4DYukawa}). 
With the $U(2)$ imposed $f^{(0)}(c_{d2})$=$f^{(0)}(c_{d1})$, the above expression dictates that even if the third generation is much heavier than the first two, 
 the effect on flavour violation due to third generation does not decouple. This feature is not restricted to Little RS but applies to most New Physics models \cite{Barbieri:2011ci,Barbieri:2015yvd}.

\section{Summary and Outlook}

Randall Sundrum models offer one of the elegant solutions not only to the hierarchy problem but also to the fermion mass hierarchies and mixing in terms of a geometric Froggatt-Nielsen mechanism. However the constraints from LEP are very stringent on these models restricting
its `visibility' at the LHC. The Little RS offers hope in that direction due to its lowered UV scale and thus larger couplings and overlap functions compared to the RS. But for the same reason, it also suffers stronger constraints from flavour physics compared to the RS. In this work we looked at the Kaon sector which has been emphasised in \cite{Bauer:2008xb} and tried to propose two solutions that help in alleviating the concerns. The first one is the use of Brane Localised Kinetic Terms for the gluon wavefunction. The second one is to use of Minimal Flavour Protection flavour symmetries to restrict certain dominant operators. For large enough BLKTs we have seen that the constraints \cite{Santiago:2008vq} reduce significantly. In the case of MFP the imposition of U(3)
works  better compared to U(2). While both these mechanisms work very differently, they are both efficient in reducing the constraints. 

Phenomenologically we expect both these mechanisms to have significantly different implications especially for electroweak precision observables and LHC signatures. Before we close two comments are in order: (a) It would be definitely interesting to see the implications of the BLKTs and flavour symmetries on systems other than $s-d$ like $b-d$ or $b-s$. In some cases, BLKTs on the gluon field might not be sufficient. (b) There could be other mechanisms like Minimal Flavour Violation (MFV) \cite{Ambrosio} which could as well suppress the New Physics contributions in a similar manner as MFP \cite{Randallperez}. This would be interesting to explore as well. 

Another line of research testing the  $b-s$-transitions is the B-anomalies \cite{gino}.
Several models, leptoquarks, $Z'$, ..,  with New Physics scale of few TeV have been constructed in order to explain the so called B-anomalies~\cite{1821789,Altmannshofer:2014cfa,Crivellin:2015lwa,Crivellin:2016ekz,Buttazzo:2017ixm,gudrun}.
 If  B-anomalies are confirmed, due to the low scale of New Physics required, this will be kind of a revolution compared to the traditional paradigm of Minimal Flavour Violation where a gap is expected between the scale of the solution of the Naturalness problem and the MFV scale~\cite{gino}.
Our scenarios depart from MFV and seem suitable to address B-anomalies: we plan to pursue the set-ups studied here by looking for flavour signatures.

\section*{Acknowledgements}
A.K. is supported by the INFN research initiative Exploring New Physics (ENP). M.T.A. acknowledges the support from SERB National Postdoctoral fellowship [PDF/2017/001350] and UGC-DSKPDF. G.D. was supported in part by MIUR under Project No. 2015P5SBHT and by the INFN research initiative ENP. G.D. thanks “Satish Dhawan Visiting Chair Professorship”  at the Indian Institute of Science. We thank IoE funds of IISc for support. SKV is supported by the project "Nature of New Physics" by the Department of Science and Technology, Govt. of India. We thank Andreas Crivellin for comments on the manuscript. 
\appendix
\section{Flavour Parameters and Numerical Analysis}\label{FlavouParameters} 
In this section we analyze the  independent flavour parameters in the quark sector and discuss the full parameter scan of c values of quark, which could fit the CKM matrices and quark masses under experimental uncertainties.\\
To understand the flavour parameters in RS model, we follow the approach of Ref.~\cite{Agashe:2004cp}. In our model we got $3\times3$ complex matrices of 5-D Yukawa couplings $Y_{{u,d}}^{(5)}$, each contains 9 real and 9 complex parameters. In RS model we get additional flavour parameters through $3\times3$ Hermitian bulk mass matrices, $c_{Q,u,d}$. This brings in additional 18 real parameters and 9 complex phases.\\
For Numerical scan it is convenient to work in the  basis where the bulk mass matrices $c_{Q,u,d}$ are diagonal and comprise of 9 real parameters.
The remaining 18 real parameters and 10 physical phases are then collected in the 5D
Yukawa coupling matrices $Y_{{u,d}}^{(5)}$.\\
We use the parameterisation adopted in the~\cite{Albrecht:2009xr}, where the bulk mass matrices are real and diagonal. We derive the parameterisation of the unitary matrices $U_{l,r}$ and $D_{l,r}$ generalizing the usual CKM parameterisation, as product of three rotations and introducing a complex phases in each of them ~\cite{Casas:2001sr}, this parameterisation ensures the information about only the physical parameters. 
We therefore aim to derive a parameterisation of the RS flavour sector in terms of the SM quark masses, the CKM parameters, and the parameters of the new flavour mixing matrices ${D}_l$, ${U}_r$ and ${D}_r$.
Neglecting the mixing with fermionic KK modes and approximating the Higgs field to be exactly localised on the IR brane, we can write
\begin{eqnarray}
\xi_{u} &=& {U}_l^{\dagger} Y_{u} {U}_r\,,\\
\xi_{d} &=& {D}_l^{\dagger} Y_{d} {D}_r\
\end{eqnarray}
where $Y_{u,d}$ are effective 4D couplings defined in Eq.(\ref{Effective4DYukawa}) and $\xi_{u,d}$ are defined as
\begin{equation}
\xi_{u}=\frac{\sqrt{2}}{v}\text{diag}(m_u,m_c,m_t),~~~~\xi_{d}=\frac{\sqrt{2}}{v}\text{diag}(m_d,m_s,m_b)  
\end{equation}
The CKM matrix is given by 
\be
{ {U}}_l^\dagger{{D}}_l=V_\text{CKM}\,.
\ee
The 5D Yukawa couplings can then be written as
\begin{eqnarray}
\left(Y_{u}^{(5)}\right)_{ij} &=&   f^{(0)-1}(c_{Qi}) \left({U}_l \xi_{u}   {U}_r^\dagger\right)_{ij} f^{(0)-1}(c_{uj})\nonumber\\
&=& f^{(0)-1}(c_{Qi}) \left({D}_l V_{CKM}^{\dagger} \xi_{u}   {U}_r^\dagger \right)_{ij} f^{(0)-1}(c_{uj}),\label{eq:lambdau}\\
\left(Y_{d}^{(5)}\right)_{ij} &=& f^{(0)-1}(c_{Qi}) \left({U}_l \xi_{u}   {U}_r^\dagger\right)_{ij} f^{(0)-1}(c_{dj})\label{eq:lambdad}\,,
\end{eqnarray}
In the above parametrization, the SM parameters are encoded in the quark masses and $V_\text{CKM}$. 9 new real flavour parameters are present in $c_{Q,u,d}$. The remaining 9 real parameters and 9 complex phases are distributed among ${D}_l$, ${U}_r$ and ${D}_r$.\\
We use CKM parameterisation, as a product of three rotations, and introducing a complex
phase in each of them, thus obtaining
product of three rotation matrices with angle $\omega_{ij}^{D_l}$ ($i,j=1,2,3$) a complex phase $\tau_{ij}^{D_l}$ ($i,j=1,2,3$) in each of them \cite{Blanke:2006xr}, Defining
\begin{equation}
c_{ij}^{D_l} = \cos\omega_{ij}^{D_l}\,,\qquad
s_{ij}^{D_l} = \sin\omega_{ij}^{D_l}\qquad (i,j=1,2,3)\,.
\end{equation}
	parameterisation of ${D}_l$ reads \cite{Blanke:2006xr}
\begin{eqnarray}
{D}_l = 
\footnotesize
\begin{pmatrix} 
c_{12}^{D_l} c_{13}^{D_l} & s_{12}^{D_l} c_{13}^{D_l} e^{-i\tau^{D_l}_{12}}& s_{13}^{D_l} e^{-i\tau^{D_l}_{13}}\\
-s_{12}^{D_l} c_{23}^{D_l} e^{i\tau^{D_l}_{12}}-c_{12}^{D_l} s_{23}^{D_l}s_{13}^{D_l} e^{i(\tau^{D_l}_{13}-\tau^{D_l}_{23})} &
c_{12}^{D_l} c_{23}^{D_l}-s_{12}^{D_l} s_{23}^{D_l}s_{13}^{D_l} e^{i(\tau^{D_l}_{13}-\tau^{D_l}_{12}-\tau^{D_l}_{23})} &
s_{23}^{D_l}c_{13}^{D_l} e^{-i\tau^{D_l}_{23}}\\
s_{12}^{D_l} s_{23}^{D_l} e^{i(\tau^{D_l}_{12}+\tau^{D_l}_{23})}-c_{12}^{D_l} c_{23}^{D_l}s_{13}^{D_l} e^{i\tau^{D_l}_{13}} &
-c_{12}^{D_l} s_{23}^{D_l} e^{i\tau^{D_l}_{23}}-s_{12}^{D_l} c_{23}^{D_l}s_{13}^{D_l} e^{i(\tau^{D_l}_{13}-\tau^{D_l}_{12})} &
c_{23}^{D_l}c_{13}^{D_l}
\end{pmatrix}
\end{eqnarray}
\normalsize
$ {U}_r$ and $ {D}_r$ are written in a completely analogous way. 
In order to naturally obtain anarchic 5D Yukawa matrices $\left(Y_{u,d}^{(5)}\right)$, it {can} in practice be useful to adapt the above parameterisation. 
The phases $\tau_{ij}^{D_l}$,  
$\tau_{ij}^{{U}_r}$ and 
$\tau_{ij}^{D_r}$
are all chosen to lie in their natural range $0\le\tau<2\pi$, however the case of the mixing angles 
$\omega_{ij}^{D_l}$, 
$\omega_{ij}^{{U}_r}$ and $
\omega_{ij}^{D_r}$
is somewhat different. Here one finds \cite{Agashe:2004cp} that anarchic 5D Yukawa couplings imply the hierarchies
\begin{equation}
\omega_{ij}^{D_l}\sim \frac{f^{(0)}(c_{Qi})}{f^{(0)}(c_{Qj})}\,,\qquad
\omega_{ij}^{{U}_r}\sim \frac{f^{(0)}(c_{ui})}{f^{(0)}(c_{uj})}
\,,\qquad
\omega_{ij}^{D_r}\sim \frac{f^{(0)}(c_{di})}{f^{(0)}(c_{dj})}\,.
\end{equation}
We can now use this knowledge to find a parameterisation that automatically leads to a natural structure for $Y^{(5)}_{u,d}$. Therefore we define
\begin{equation}
\omega_{ij}^{D_l}= \epsilon_{ij}^{D_l} \frac{f^{(0)}(c_{Qi})}{f^{(0)}(c_{Qj})}\,,\qquad
\omega_{ij}^{{U}_r} =  \epsilon_{ij}^{{U}_r}\frac{f^{(0)}(c_{ui})}{f^{(0)}(c_{uj})}
\,,\qquad
\omega_{ij}^{D_r}=  \epsilon_{ij}^{D_r} \frac{f^{(0)}(c_{di})}{f^{(0)}(c_{dj})}\,,
\end{equation}
where $\epsilon_{ij}^{D_l}$, $\epsilon_{ij}^{{U}_r}$, $\epsilon_{ij}^{D_r}$ are $\ord(1)$ parameters. 
We choose to scan over the following independent variables $c_{Q,u,d}$, $U_{l,r}$ and $D_{l,r}$, then we check whether the 5D Yukawa is anarchic and satisfy the condition $0.1\le|Y_{{u,d}_{ij}}^{(5)}|\le3$. 
\begin{table}[h]
	\begin{center}
		\def\arraystretch{0.7}
		\begin{tabular}{|c|c|}
			\hline
			Parameter & Value $\pm$ Error \\[0.2mm] 
			\hline 
			$m_t$ & $(136.2 \pm 3.1) \, {\rm GeV}$  \\
			$m_b$ & $(2.4 \pm 0.04) \, {\rm GeV}$ \\
			$m_c$ & $(0.56 \pm 0.04) \, {\rm GeV}$  \\
			$m_s$ & $(0.047 \pm 0.012) \, {\rm MeV}$  \\
			$m_d$ & $(2.0 \pm 4.0) \, {\rm MeV}$  \\
			$m_u$ & $(1.2 \pm 0.4) \, {\rm MeV}$ \\
			\hline 
		\end{tabular}
	\end{center}
	\caption{Quark  $\overline{MS}$ masses at 3 TeV \cite{Csaki:2008zd}.}\label{Quarkmass}
\label{fermionmass}
\end{table}
The rough choice of c values of fermions
for anarchic 5-D Yukawa couplings are approximated using rough size of
the mixing angles given by~\cite{Huber:2003tu}:
\begin{eqnarray}
|D_{l}|_{ij} \sim \frac{f^{(0)}(c_{Qi})}{f^{(0)}(c_{Qj})}\,,\qquad
|D_{r}|_{ij} \sim \frac{f^{(0)}(c_{di})}{f^{(0)}(c_{dj})}\,,\qquad
|U_{r}|_{ij} \sim \frac{f^{(0)}(c_{ui})}{f^{(0)}(c_{uj})}\,,\quad
i\lessgtr j
\end{eqnarray}  
We can also get that $|V_{CKM}|_{ij}\sim\frac{f^{(0)}(c_{Qi})}{f^{(0)}(c_{Qj})}$, thus the hierarchy in CKM elements is purely set by the $c_{Q}$. The hierarchy in the mass eigenvalues are given as:
\begin{eqnarray}
(m_{u,d})_{i}\sim \frac{v}{\sqrt{2}}Yf^{(0)}(c_{Qi})f^{(0)}(c_{u,di})
\end{eqnarray}
The CKM and the mass eigenvalues fixes the $f^{(0)}(c_{Q,u,d})$ in terms of CKM parameters and fermions masses as (assuming $f^{(0)}(c_{u3})\sim \ord{(1)}$):
\begin{eqnarray}
f^{(0)}(c_{Q2})/f^{(0)}(c_{Q3})\sim \lambda^{2}\,,\quad
f^{(0)}(c_{Q1})/f^{(0)}(c_{Q3})\sim \lambda^{3}\,,\quad
f^{(0)}(c_{u1})\sim \frac{m_u}{m_t}\frac{1}{\lambda^{3}}\,,\quad
f^{(0)}(c_{u2})\sim \frac{m_c}{m_t}\frac{1}{\lambda^{2}}\,,\quad
\label{eq:up}
\end{eqnarray}
\begin{eqnarray}
f^{(0)}(c_{d1})\sim \frac{m_d}{m_t}\frac{1}{\lambda^{3}}\,,\quad
f^{(0)}(c_{d2})\sim\frac{m_s}{m_t}\frac{1}{\lambda^{2}}\,,\quad
f^{(0)}(c_{d3})\sim \frac{m_b}{m_t}\,\quad
\label{eq:para1}
\end{eqnarray} 
where $\lambda \sim 0.22$ .
For simplicity of numerical analysis, we randomly vary each set of the independent $c$ values around their natural size Eq.(\ref{eq:up}),Eq.(\ref{eq:para1}) by a factor of three. The diagonalization matrix is calculated using the above mentioned technique. The masses of quarks used in this work is given in Table \ref{Quarkmass}. We choose the $c_{Q3}$ from Ref.\cite{Bauer:2009cf, Casagrande:2008hr} which  allows for consistency with electro-weak precision data as to satisfy the $Z\rightarrow b_{l}\bar{b_{l}}$.
\bibliographystyle{ieeetr}
\bibliography{Notes.bib}

\begin{thebibliography}{10}

\bibitem{Beall:1981ze}
G.~Beall, M.~Bander, and A.~Soni, ``{Constraint on the Mass Scale of a
  Left-Right Symmetric Electroweak Theory from the K(L) K(S) Mass
  Difference},'' {\em Phys. Rev. Lett.}, vol.~48, p.~848, 1982.

\bibitem{Buchalla:1995vs}
G.~Buchalla, A.~J. Buras, and M.~E. Lautenbacher, ``{Weak decays beyond leading
  logarithms},'' {\em Rev. Mod. Phys.}, vol.~68, pp.~1125--1144, 1996.

\bibitem{Buras:1998raa}
A.~J. Buras, ``{Weak Hamiltonian, CP violation and rare decays},'' in {\em {Les
  Houches Summer School in Theoretical Physics, Session 68: Probing the
  Standard Model of Particle Interactions}}, pp.~281--539, 6 1998.

\bibitem{Isidori:2010kg}
G.~Isidori, Y.~Nir, and G.~Perez, ``{Flavor Physics Constraints for Physics
  Beyond the Standard Model},'' {\em Ann. Rev. Nucl. Part. Sci.}, vol.~60,
  p.~355, 2010.

\bibitem{Kiers:2002cz}
K.~Kiers, J.~Kolb, J.~Lee, A.~Soni, and G.-H. Wu, ``{Ubiquitous CP violation in
  a top inspired left-right model},'' {\em Phys. Rev. D}, vol.~66, p.~095002,
  2002.

\bibitem{Gori:2016lga}
S.~Gori, ``{Three Lectures of Flavor and CP violation within and Beyond the
  Standard Model},'' in {\em {2015 European School of High-Energy Physics}},
  pp.~65--90, 2017.

\bibitem{Zupan:2019uoi}
J.~Zupan, ``{Introduction to flavour physics},'' {\em CERN Yellow Rep. School
  Proc.}, vol.~6, pp.~181--212, 2019.

\bibitem{Randall_1999_1}
L.~Randall and R.~Sundrum, ``An alternative to compactification,'' {\em
  Physical Review Letters}, vol.~83, p.~4690–4693, Dec 1999.

\bibitem{Randall_1999_2}
L.~Randall and R.~Sundrum, ``Large mass hierarchy from a small extra
  dimension,'' {\em Physical Review Letters}, vol.~83, p.~3370–3373, Oct
  1999.

\bibitem{Agashe:2003zs}
K.~Agashe, A.~Delgado, M.~J. May, and R.~Sundrum, ``{RS1, custodial isospin and
  precision tests},'' {\em JHEP}, vol.~08, p.~050, 2003.

\bibitem{Agashe:2004cp}
K.~Agashe, G.~Perez, and A.~Soni, ``{Flavor structure of warped extra dimension
  models},'' {\em Phys. Rev.}, vol.~D71, p.~016002, 2005.

\bibitem{Davoudiasl:2008hx}
H.~Davoudiasl, G.~Perez, and A.~Soni, ``{The Little Randall-Sundrum Model at
  the Large Hadron Collider},'' {\em Phys. Lett.}, vol.~B665, pp.~67--71, 2008.

\bibitem{Davoudiasl_2010}
H.~Davoudiasl, S.~Gopalakrishna, and A.~Soni, ``Big signals of little
  randall–sundrum models,'' {\em Physics Letters B}, vol.~686, p.~239–243,
  Mar 2010.

\bibitem{Bauer:2008xb}
M.~Bauer, S.~Casagrande, L.~Grunder, U.~Haisch, and M.~Neubert, ``{Little
  Randall-Sundrum models: epsilon(K) strikes again},'' {\em Phys. Rev. D},
  vol.~79, p.~076001, 2009.

\bibitem{Dvali:2000hr}
G.~Dvali, G.~Gabadadze, and M.~Porrati, ``{4-D gravity on a brane in 5-D
  Minkowski space},'' {\em Phys. Lett. B}, vol.~485, pp.~208--214, 2000.

\bibitem{Georgi:2000ks}
H.~Georgi, A.~K. Grant, and G.~Hailu, ``{Brane couplings from bulk loops},''
  {\em Phys. Lett. B}, vol.~506, pp.~207--214, 2001.

\bibitem{Carena:2002dz}
M.~Carena, E.~Ponton, T.~M.~P. Tait, and C.~E.~M. Wagner, ``{Opaque branes in
  warped backgrounds},'' {\em Phys. Rev.}, vol.~D67, p.~096006, 2003.

\bibitem{Carena:2003fx}
M.~Carena, A.~Delgado, E.~Ponton, T.~M. Tait, and C.~Wagner, ``{Precision
  electroweak data and unification of couplings in warped extra dimensions},''
  {\em Phys. Rev. D}, vol.~68, p.~035010, 2003.

\bibitem{Santiago:2008vq}
J.~Santiago, ``{Minimal Flavor Protection: A New Flavor Paradigm in Warped
  Models},'' {\em JHEP}, vol.~12, p.~046, 2008.

\bibitem{Gherghetta:2000qt}
T.~Gherghetta and A.~Pomarol, ``{Bulk fields and supersymmetry in a slice of
  AdS},'' {\em Nucl. Phys. B}, vol.~586, pp.~141--162, 2000.

\bibitem{Casagrande:2008hr}
S.~Casagrande, F.~Goertz, U.~Haisch, M.~Neubert, and T.~Pfoh, ``{Flavor Physics
  in the Randall-Sundrum Model: I. Theoretical Setup and Electroweak Precision
  Tests},'' {\em JHEP}, vol.~10, p.~094, 2008.

\bibitem{Iyer:2015ywa}
A.~M. Iyer, K.~Sridhar, and S.~K. Vempati, ``{Bulk Randall-Sundrum models,
  electroweak precision tests, and the 125 GeV Higgs},'' {\em Phys. Rev. D},
  vol.~93, no.~7, p.~075008, 2016.

\bibitem{Aaboud_2018}
M.~Aaboud, G.~Aad, B.~Abbott, O.~Abdinov, B.~Abeloos, S.~H. Abidi, O.~S.
  AbouZeid, N.~L. Abraham, H.~Abramowicz, and et~al., ``Search for heavy
  particles decaying into top-quark pairs using lepton-plus-jets events in
  proton–proton collisions at $\sqrt{s} = 13 \text {TeV}$ with the atlas
  detector,'' {\em The European Physical Journal C}, vol.~78, Jul 2018.

\bibitem{calchep}
A.~Belyaev, N.~D. Christensen, and A.~Pukhov, ``{CalcHEP 3.4 for collider
  physics within and beyond the Standard Model},'' {\em Comput. Phys. Commun.},
  vol.~184, pp.~1729--1769, 2013.

\bibitem{Csaki:2008zd}
C.~Csaki, A.~Falkowski, and A.~Weiler, ``{The Flavor of the Composite
  Pseudo-Goldstone Higgs},'' {\em JHEP}, vol.~09, p.~008, 2008.

\bibitem{Bauer:2009cf}
M.~Bauer, S.~Casagrande, U.~Haisch, and M.~Neubert, ``{Flavor Physics in the
  Randall-Sundrum Model: II. Tree-Level Weak-Interaction Processes},'' {\em
  JHEP}, vol.~09, p.~017, 2010.

\bibitem{Agashe:2008uz}
K.~Agashe, A.~Azatov, and L.~Zhu, ``{Flavor Violation Tests of Warped/Composite
  SM in the Two-Site Approach},'' {\em Phys. Rev. D}, vol.~79, p.~056006, 2009.

\bibitem{Blanke:2008zb}
M.~Blanke, A.~J. Buras, B.~Duling, S.~Gori, and A.~Weiler, ``{$\Delta$ F=2
  Observables and Fine-Tuning in a Warped Extra Dimension with Custodial
  Protection},'' {\em JHEP}, vol.~03, p.~001, 2009.

\bibitem{Ahmed:2019zxm}
A.~Ahmed, A.~Carmona, J.~Castellano~Ruiz, Y.~Chung, and M.~Neubert,
  ``{Dynamical origin of fermion bulk masses in a warped extra dimension},''
  {\em JHEP}, vol.~08, p.~045, 2019.

\bibitem{Ciuchini:1998ix}
M.~Ciuchini {\em et~al.}, ``{Delta M(K) and epsilon(K) in SUSY at the
  next-to-leading order},'' {\em JHEP}, vol.~10, p.~008, 1998.

\bibitem{Ciuchini:1997bw}
M.~Ciuchini, E.~Franco, V.~Lubicz, G.~Martinelli, I.~Scimemi, and
  L.~Silvestrini, ``{Next-to-leading order QCD corrections to Delta F = 2
  effective Hamiltonians},'' {\em Nucl. Phys.}, vol.~B523, pp.~501--525, 1998.

\bibitem{Bagger_1997}
J.~A. Bagger, K.~T. Matchev, and R.-J. Zhang, ``Qcd corrections to
  flavor-changing neutral currents in the supersymmetric standard model,'' {\em
  Physics Letters B}, vol.~412, p.~77–85, Oct 1997.

\bibitem{Buras_2000}
A.~J. Buras, M.~Misiak, and J.~Urban, ``Two-loop qcd anomalous dimensions of
  flavour-changing four-quark operators within and beyond the standard model,''
  {\em Nuclear Physics B}, vol.~586, p.~397–426, Oct 2000.

\bibitem{Barbieri:2011ci}
R.~Barbieri, G.~Isidori, J.~Jones-Perez, P.~Lodone, and D.~M. Straub, ``{$U(2)$
  and Minimal Flavour Violation in Supersymmetry},'' {\em Eur. Phys. J. C},
  vol.~71, p.~1725, 2011.

\bibitem{Barbieri:2015yvd}
R.~Barbieri, G.~Isidori, A.~Pattori, and F.~Senia, ``{Anomalies in $B$-decays
  and $U(2)$ flavour symmetry},'' {\em Eur. Phys. J. C}, vol.~76, no.~2, p.~67,
  2016.

\bibitem{Ambrosio}
G.~D'Ambrosio, G.~Giudice, G.~Isidori, and A.~Strumia, ``{Minimal flavor
  violation: An Effective field theory approach},'' {\em Nucl. Phys. B},
  vol.~645, pp.~155--187, 2002.

\bibitem{Randallperez}
A.~Fitzpatrick, G.~Perez, and L.~Randall, ``{Flavor anarchy in a
  Randall-Sundrum model with 5D minimal flavor violation and a low Kaluza-Klein
  scale},'' {\em Phys. Rev. Lett.}, vol.~100, p.~171604, 2008.

\bibitem{gino}
G.~Isidori, ``Model building on flavour anomalies and implications for high-pt
  searches at implications of lhcb measurements and future prospects, cern
  geneva,'' Nov 2017.

\bibitem{1821789}
M.~Bordone, O.~Cata, T.~Feldmann, and R.~Mandal, ``{Constraining flavour
  patterns of scalar leptoquarks in the effective field theory},'' 10 2020.

\bibitem{Altmannshofer:2014cfa}
W.~Altmannshofer, S.~Gori, M.~Pospelov, and I.~Yavin, ``{Quark flavor
  transitions in $L_\mu-L_\tau$ models},'' {\em Phys. Rev. D}, vol.~89,
  p.~095033, 2014.

\bibitem{Crivellin:2015lwa}
A.~Crivellin, G.~D'Ambrosio, and J.~Heeck, ``{Addressing the LHC flavor
  anomalies with horizontal gauge symmetries},'' {\em Phys. Rev. D}, vol.~91,
  no.~7, p.~075006, 2015.

\bibitem{Crivellin:2016ekz}
A.~Crivellin, ``{B-anomalies related to leptons and lepton flavour universality
  violation},'' {\em PoS}, vol.~BEAUTY2016, p.~042, 2016.

\bibitem{Buttazzo:2017ixm}
D.~Buttazzo, A.~Greljo, G.~Isidori, and D.~Marzocca, ``{B-physics anomalies: a
  guide to combined explanations},'' {\em JHEP}, vol.~11, p.~044, 2017.

\bibitem{gudrun}
G.~Hiller, ``Flavor theory 2020 and outlook having fun with leptons:
  B-anomalies and more, ichep2020 prague,''

\bibitem{Albrecht:2009xr}
M.~E. Albrecht, M.~Blanke, A.~J. Buras, B.~Duling, and K.~Gemmler,
  ``{Electroweak and Flavour Structure of a Warped Extra Dimension with
  Custodial Protection},'' {\em JHEP}, vol.~09, p.~064, 2009.

\bibitem{Casas:2001sr}
J.~A. Casas and A.~Ibarra, ``{Oscillating neutrinos and muon ---> e, gamma},''
  {\em Nucl. Phys.}, vol.~B618, pp.~171--204, 2001.

\bibitem{Blanke:2006xr}
M.~Blanke, A.~J. Buras, A.~Poschenrieder, S.~Recksiegel, C.~Tarantino,
  S.~Uhlig, and A.~Weiler, ``{Another look at the flavour structure of the
  littlest Higgs model with T-parity},'' {\em Phys. Lett.}, vol.~B646,
  pp.~253--257, 2007.

\bibitem{Huber:2003tu}
S.~J. Huber, ``{Flavor violation and warped geometry},'' {\em Nucl. Phys.},
  vol.~B666, pp.~269--288, 2003.

\end{thebibliography}
\end{document}